 \documentclass[final,5p,times,twocolumn]{elsarticle}

\usepackage{amsmath}
\usepackage{booktabs}
\usepackage{subfigure}
\usepackage{amssymb}
\usepackage{amsthm}
\usepackage{threeparttable }
\usepackage{float}
\usepackage{algorithm}
\usepackage{algorithmicx}
\usepackage{algpseudocode}

\newcommand{\brho}{\mbox{\boldmath$\rho$}}

\def\f{\mathbf{f}}
\def\c{C}

\def\U{\mathcal{U}}
\def\M{\mathcal{M}}
\def\H{\mathcal{H}}

\def\u{\mathbf{u}}

\def\K{\mathbf{K}}
\def\RR{\mathbb{R}}

\def\eb{\begin{equation}}
\def\ee{\end{equation}}

\begin{document}

\begin{frontmatter}



\title{Self-supporting Topology Optimization for Additive Manufacturing}




\author[cadaddress]{Dengyang Zhao}
\author[cadaddress]{Ming Li\corref{mycorrespondingauthor}}
\author[cadaddress]{Yusheng Liu}
\cortext[mycorrespondingauthor]{Corresponding author: liming@cad.zju.edu.cn}
\address[cadaddress]{State Key Laboratory of CAD\&CG, Zhejiang University, Hangzhou, China}

\begin{abstract}
The paper presents a topology optimization approach that designs an optimal structure, called a \emph{self-supporting structure}, which is ready to be fabricated via additive manufacturing without the usage of additional \emph{support} structures. Such supports in general have to be created during the fabricating process so that the primary object can be manufactured layer by layer without collapse, which is very time-consuming and waste of material.

The proposed approach resolves this problem by formulating the self-supporting requirements as a novel explicit quadratic continuous constraint in the topology optimization problem, or specifically, requiring the number of unsupported elements (in terms of the sum of squares of their densities) to be zero. Benefiting form such novel formulations, computing sensitivity of the self-supporting constraint with respect to the design density is straightforward, which otherwise would require lots of research efforts in general topology optimization studies. The derived sensitivity for each element is only linearly dependent on its sole density, which, different from previous layer-based sensitivities, consequently allows for a parallel implementation and possible higher convergence rate. In addition, a discrete convolution operator is also designed to detect the unsupported elements as involved in each step of optimization iteration, and improves the detection process 100 times as compared with simply enumerating these elements. The approach works for cases of general overhang angle, or general domain, and produces an optimized structures, and their associated optimal compliance, very close to that of the reference structure obtained without considering the self-supporting constraint, as demonstrated by extensive 2D and 3D benchmark examples.
\end{abstract}


\end{frontmatter}

\section{Introduction}
Topology optimization aims to generate an optimal material distribution within a design domain under certain geometric or physical constraints. Since its introduction in late the 1980s~\cite{Bendsoe88}, this problem has attracted wide industrial and academic interests due to its large potentiality in engineering applications and its intrinsic mathematical challenges. Topology optimization has developed in many different forms, such as: homogenization~\cite{Bendsoe88}, density (SIMP)~\cite{Bendsoe89}, evolutionary approaches (BESO)~\cite{xie1993simple,Huang2010}, level set~\cite{Wang2003,VanDijk2013}, or more recently IGA (iso-geometric analysis)~\cite{Qian2013}, to name a few. See~\cite{Sigmund2013} for a recent and comprehensive review on this topic.

The complex geometric designs produced by topology optimization show the approach's superiority in balancing the geometric distribution and the target physical performance. Such designs are however very difficult to be manufactured directly via traditional subtractive or formative manufacturing techniques~\cite{Sigmund2006On,Serphos2014Incorporating,Liu2016}. On the other hand, rapidly developing additive manufacturing technologies have the promise to overcome the barrier between the potentiality that the topology optimization approaches can provide and the limitations that traditional manufacturing technologies can fabricate. In reality, additive manufacturing is a natural counterpart to topology optimization in that they have very versatile capability to quickly generate and realize new components not existing before~\cite{Lang20161,Lang20162}. 

Despite the enhanced geometric freedom associated with additive manufacturing, specific design rules must still be satisfied in order to ensure manufacturability. The fabrication overhang angle is an example of a rule which is of paramount importance to ensure that the part will not collapse when fabricating the designed structure layer by layer. A structure satisfying such an overhang angle constraint is called \emph{self-supporting}. For example, Thomas~\cite{Thomas2010} identified 45 degree as the typical maximum overhang angle with a large number of experiments. For a non self-supporting structure, its geometry has to be modified or additional \emph{support} structures need to be generated. Modifying the geometry will ultimately reduce the structure's physical performance, while additional support raises the issue of automatic and minimum volume support design~\cite{Majhi1999,Vanek2014Clever,Wang2013,Dumas2014Bridging,Wu2016}, and further post-processing activities to remove the unwanted supports. In the case that the support is made of the same material as the main component, such as the selective laser melting (SLM) process using metals, it is extremely difficult to remove out the support structure. Particularly, when the generated supports are embedded within a closed volume of the model, it is impossible to remove them.

The best strategy to resolve the issue of topology optimization for additive manufacturing is perhaps to design a completely self-supporting structure, via topology optimization, that can be fabricated directly without the usage of support materials. Brackett et al first suggested including the overhang angle constraints into the topology optimization process~\cite{Brackett2011topology}, but does not produce a complete self-supporting structure. The first self-supporting structure built from topology optimization is due to the pioneering work of Gaynor and Guest in 2014~\cite{Gaynor2014} (and a very recent journal version~\cite{Gaynor2016}), which is achieved via introducing a wedge-shaped filter during the topology optimization process. Also very recently, an excellent work was also conducted by Langelaar in 2D~\cite{Lang20161} and in 3D~\cite{Lang20162} via introducing a novel self-supporting filter into the topology optimization process, which is achieved via building smooth approximation to the minimum and maximum functions. Impressive 2D and 3D examples were also shown in these studies~\cite{Gaynor2016,Lang20161,Lang20162}.

In this paper, an alternative novel self-supporting topology optimization approach is proposed to generate a structure of optimal physical performance that does not need any additional support materials. It is achieved via carefully formulating the self-supporting constraint as an \emph{explicit} quadratic function with respect to the design density, specifically, requiring the number of unsupported elements (in terms of the sum of their densities) to be zero. Benefiting from the novel quadratic formulation, the self-supporting sensitivity for each element is straightforward to compute, and is only linearly dependent on density of the element itself; notice that designing a proper filter and computing the associated sensitivity usually requires lots of research efforts in general topology optimization framework~\cite{Sigmund2013,Gaynor2016,Lang20161,Lang20162}. The derived sensitivity does not involve density information of any other elements, and thus allows for a parallel implementation, which is particularly important for 3D problem of high DOFs. Previous approaches~\cite{Gaynor2016,Lang20161,Lang20162} have a nonlinear \emph{layer-based} sensitivity expressions, and may thus inhibit parallel implementations, as also explained by the authors~\cite{Gaynor2016,Lang20161,Lang20162}. In addition, a discrete convolution operator is also designed to detect the unsupported elements as involved in each step of optimization iteration, and improves the detection process 100 times as compared with simply enumerating these elements.

Comparisons between the proposed approach and previous studies~\cite{Gaynor2014,Gaynor2016,Lang20161,Lang20162} are also summarized in Table 1.
\newcommand{\tabincell}[2]{\begin{tabular}{@{}#1@{}}#2\end{tabular}}
\begin{table*}[tb]
\caption{Difference between self-supporting topology optimization approaches}\label{tab:method-comp}
\centering
\begin{tabular}{|c|c|c|c|c|}
  \hline
  Method &\tabincell{c}{Supported Element\\Detection} &\tabincell{c}{Self-supporting \\Constraint Prescription}&\tabincell{c}{Smooth\\Approximation}&Sensitivity\\
  \hline
  \tabincell{c}{Gaynor and\\Guest~\cite{Gaynor2014,Gaynor2016}}&\tabincell{c}{Layerwise\\Enumeration}&\tabincell{c}{Area occupation}&\tabincell{c}{Heaviside\\function}&
  \tabincell{c}{Nonlinear\\Layer dependent}\\
  \hline
  \tabincell{c}{Lang~\cite{Lang20161,Lang20162}}&\tabincell{c}{Layerwise\\Enumeration}& Min-Max&\tabincell{c}{Norm\\approximation}&\tabincell{c}{Nonlinear\\Layer dependent}\\
  \hline
  \tabincell{c}{The proposed}&\tabincell{c}{Convolution}&\tabincell{c}{Quadratic\\function}&Not needed &\tabincell{c}{Linear\\Not dependent}\\
  \hline
\end{tabular}
\end{table*}

The remainder of the paper is organized as follows. The novel formulation of self-supporting topology optimization, together with its overall numerical procedure, is presented in Section~\ref{sec-problem}. Several numerical techniques behind the proposed approach are detailed in Section~\ref{sec-numerical}. Extensive 2D and 3D examples are demonstrated in Section~\ref{sec-example}. pFinally the paper is concluded in Section~\ref{sec-conclusion}.

\section{Problem statement and approach overview}\label{sec-problem}
In the section, the self-supporting constraint is formulated as a quadratic continuous function in terms of the element density, and is integrated within the classical SIMP framework~\cite{Bendsoe89} for self-supporting topology optimization. Following on from this, the proposed numerical approach to resolve the problem is outlined.

\subsection{Supported and unsupported elements}
The \emph{supported elements} generally stand for the structural elements that can be fabricated via an additive manufacturing technology without collapse with respect to the fabrication process. They are defined here using the concepts of a maximum printable supporting angle, or overhang angle, which is first assumed to be 45 degree following previous study~\cite{Brackett2011topology}. Extensions of the approach to general overhang angles are also explained later. We also assume that the printing direction is following the positive-y axis direction in both 2D and 3D for ease of explanation.

\begin{figure}[tb]
\centering
\subfigure[2D case]{\includegraphics[width=0.23\textwidth]{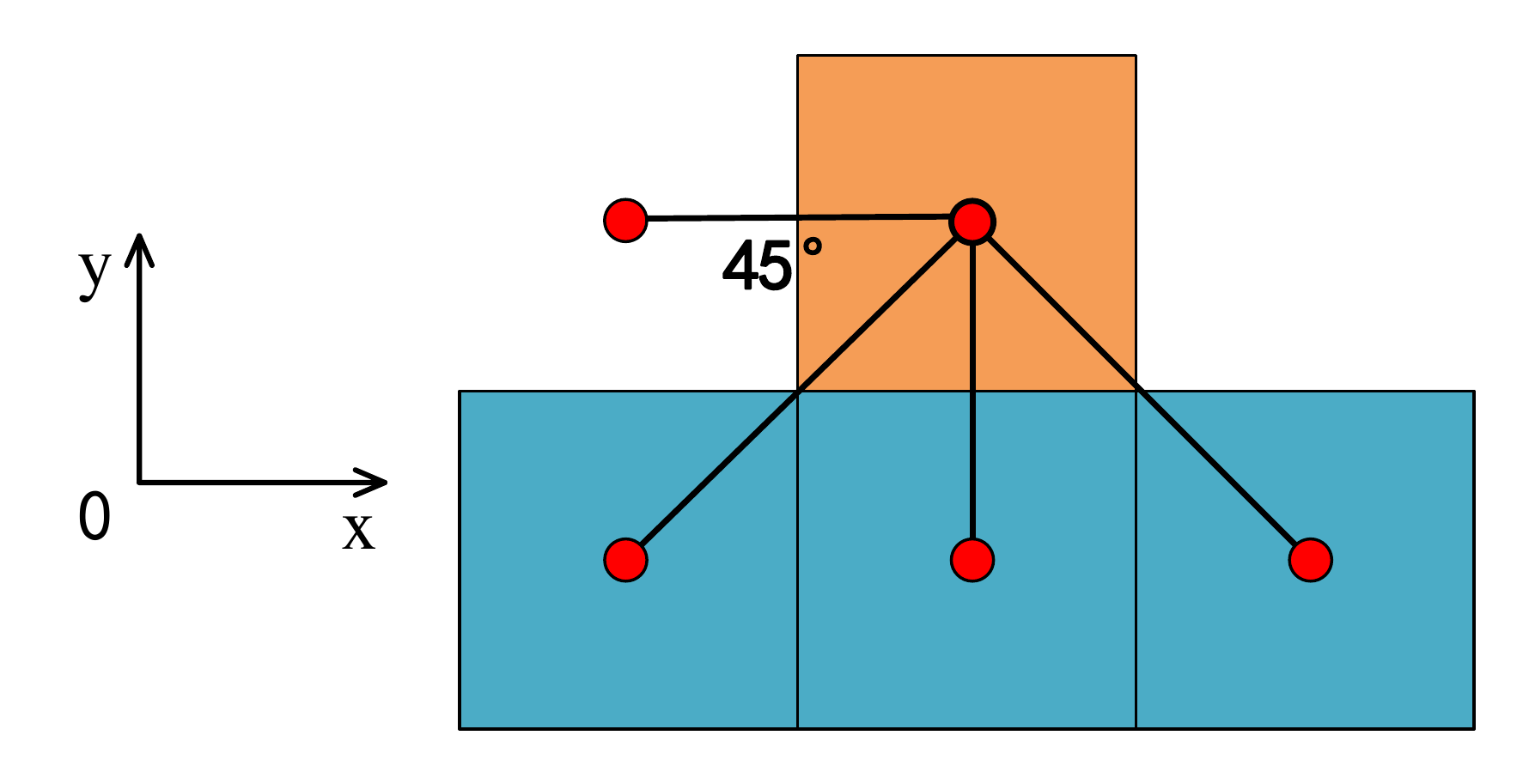}}
\subfigure[3D case]{\includegraphics[width=0.23\textwidth]{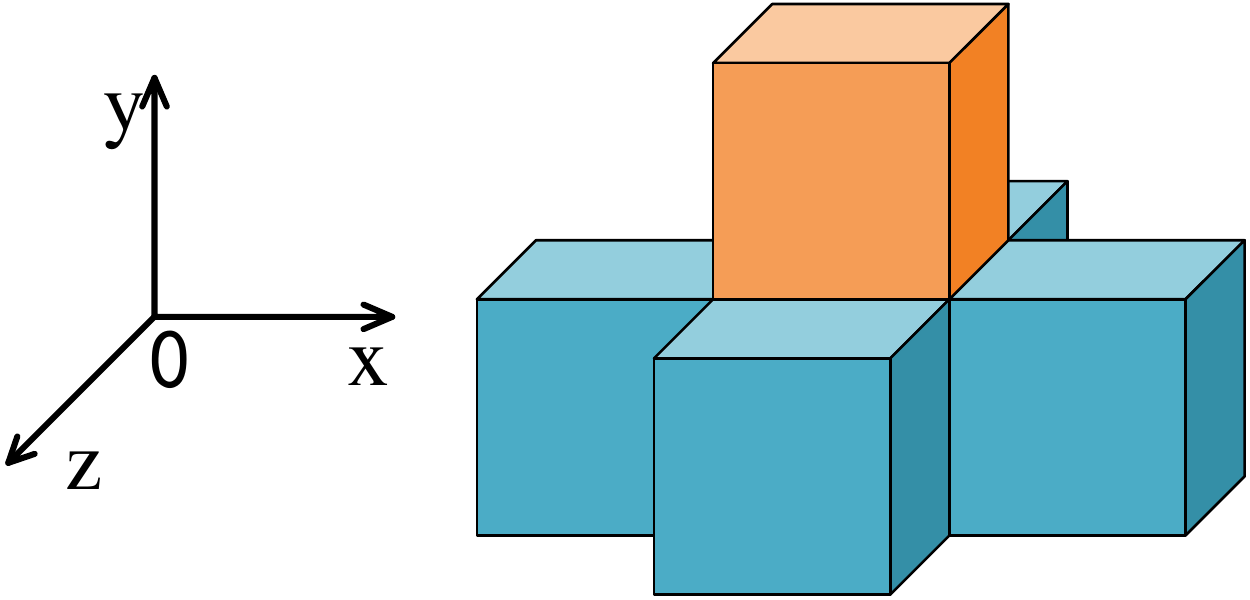}}
\caption{A supported element (in orange) is supported by one of the supporting elements in blue in 2D and 3D.}
\label{fig-23Dsp}
\end{figure}

First consider a 2D discrete structured mesh model $\M$ consisting of square elements $e(n,m)$, that is,
\[
\M=\{e(n,m)|\ 1\leq n\leq N,\ 1\leq m\leq M \},
\]
where $n,m$ are the indices increasing along the $x$ and $y$ axes respectively. Without confusion, we also use $e$ to represent a square element without explicitly mentioning its indices $n,m$. In addition, a density matrix $\brho$ of size $N\times M$ is also associated to $\M$, where an entry value $\rho(n,m)=1$ or $0$ respectively represents a solid or void element $e(n,m)$ of $\M$.

As illustrated in Fig.~\ref{fig-23Dsp}(a), given a solid element $e(n,m)$ in $\M$ (in orange), it is \emph{supported}, or called a \emph{supported element}, if one of the three blue elements below it is solid. We formulate the self-supporting condition in a continuous form as follows: an element $e(n,m)\in\M$ is supported if
\[
\sum_{n-1\leq r\leq n+1} \rho(r,m-1)> 0.
\]

Correspondingly, the \emph{supporting set} $\M_{S}$ of model $\M$ is the set of all supported elements within $\M$, that is,
\begin{equation}\label{eq-ss2D}
\M_{S}=\{e(n,m)\in\M | m=1\ \mbox{or}\ \sum_{n-1\leq r\leq n+1} \rho(r,m-1)> 0\}.
\end{equation}

Similarly, given a 3D structured mesh model $\M$ consisting of cubic elements,
\[
\M=\{e(n,m,l)|\ 1\leq n\leq N,\ 1\leq m\leq M,\ 1\leq l\leq L\},
\]
where $n,m,l$ are the indices increasing along the $x,y,z$ axes respectively, the supporting set of $\M$ is similarly defined (see also Fig.~\ref{fig-23Dsp}(b)):
\begin{equation}\label{eq-ss3D}
\M_{S}= \{e(n,m,l)\in \M | m=1\ \mbox{or} \sum_{\substack{n-1\leq r\leq n+1,\\ l-1\leq s\leq l+1}} \rho(r,m-1,s)> 0\}.
\end{equation} 
Note also here that only five elements are included here as other element do not form an appropriate overhang angle with the orange element.

Correspondingly, the set of unsupported elements of a model $\M$ is
\begin{equation}\label{eq-mu}
\M_U=\M\setminus \M_S.
\end{equation}

\subsection{Formulation of self-supporting topology optimization}
The self-supporting topology optimization problem aims to find the optimal material distribution within a design domain under certain boundary conditions. As widely studied before, the problem of minimum compliance or equivalently maximum stiffness is examined here. Following the classical SIMP framework~\cite{Bendsoe89}, the problem of self-supporting topology optimization is formulated here as an optimization problem with an additional explicit self-supporting constraint. The constraint is carefully reformulated using a simple quadratic function with respect to the density, specifically, requiring the number of unsupported elements (in terms of the sum of square of their densities) is zero. Details are explained below.

The problem of self-supporting topology optimization is stated as: find the optimal density distribution $\brho$,
\begin{equation}\label{eq-problem}
\min_{\brho\in\RR^{N\times M}}c (\u,\brho), \ s.t. \\
\end{equation}
\[
\left\{
\begin{array}{ll}
& \K( \brho)\u=\f( \brho), \\
& V(\brho)/V_0 \leq f, \\
& U(\brho) = \sum_{e \in \M_U} \rho_e^2 \leq \epsilon, \\
& 0< \rho_e\leq 1, \ e=1,\ldots, s,
\end{array}
\right.
\]
where $\brho$ is the vector of design variables (element densities) to be computed, $\u$ is the vector of global displacements and $\K$ is the global stiffness matrix. The objective function $c(\u,\brho)$ is the structure's compliance, defined as
\[
c(\u,\brho)=\u^T\K\u.
\]
$\f(\brho)$ is the nodal force vector, $V(\brho)$ and $V_0$ are the material volume and design domain volume, $f$ is the prescribed volume fraction, $\M_U$ is the index set of unsupported elements as defined in~\eqref{eq-mu}, and $\epsilon>0$ is a small parameter closed to $0$. A penalty parameter $p$, usually set as $p=3$, is applied here for the 0,1 convergence of $\brho$, or specifically,
\[
\K_e=\rho_e^p\K_e^0,
\]
where $\K_e^0$ is the element stiffness matrix associated with an element $e$ in the model $\M$ and $\rho_e$ the associated element density.

The only difference of the above conventions in~\eqref{eq-problem} with previous SIMP-based formulations is that it has an additional constraint $U(\brho) = \sum_{e \in \U} \rho_e^2 \leq \epsilon$ to meet the self-supporting requirement. This condition is based on the observation that when the sum of the element densities of unsupported elements tends to $0$, all the elements are self-supported; the square is used here so that its derivative is not constant. The simple quadratic expression allows for a straightforward sensitivity derivation of the self-supporting constraint, and ultimately results in a linear expression. 

\subsection{Approach overview}
\begin{figure*}[bt]
\centering
\includegraphics[width=0.8\textwidth]{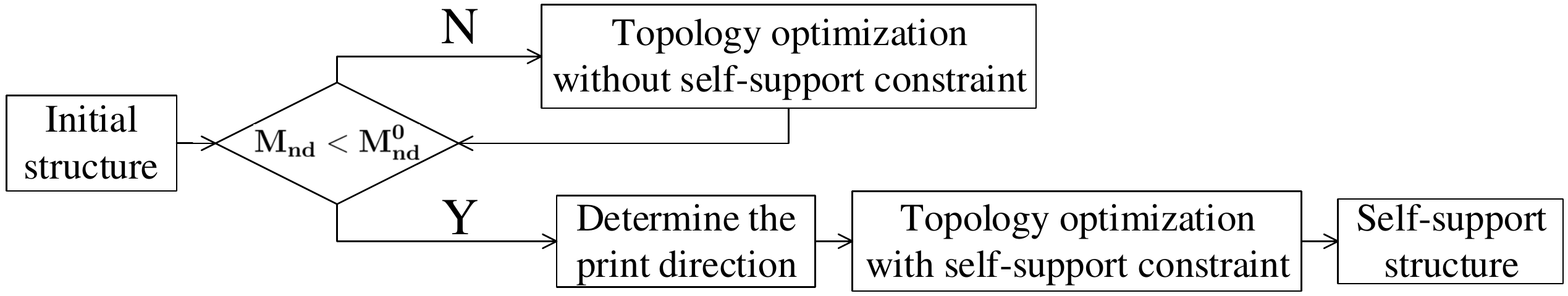}
\caption{Main procedure of the overall approach.}
\label{fig-procedure}
\end{figure*}

The self-supporting topology optimization problem~\eqref{eq-problem} is ready-to-solve using the MMA approach noticing that the sensitivity is straightforward to compute (as can be further seen in Section~\ref{sec-numerical:ss}). On the other hand, in order to further improve the approach's convergence and computational efficiency, the overall optimization process is carefully designed, as plotted in Fig.~\ref{fig-procedure} and detailed below.

Firstly, whether an element is self-supporting is dependent on the printing direction. Different printing directions produce different optimization structures. Thus, if the printing direction is chosen arbitrarily, it may produce a structure totally different from the benchmark support-needed structure obtained without considering the self-supporting constraint, with a possible worse physical performance. In some very special case, the optimization approach may not converge. Thus, an appropriate printing direction is first set via generating a coarse structure via topology optimization without considering the self-supporting constraints. It chooses the coordinate axis direction with the least number of unsupported elements as the print direction.

The criteria to generate this initial coarse structure, or to stop the initial optimization process stops, is based on the measure of non-discreteness $M_{nd}$, proposed in~\cite{Sigmund2007Morphology}. It stops when
\begin{equation}
M_{nd}<M_{nd}^0,
\end{equation}
where
\[
M_{nd}=\frac{\sum_{e=1}^N4\rho_e(1-\rho_e)}{n}\times 100\%,
\]
and $n$ is the number of elements of the domain, $\rho_e$ is the density an element $e$. The valve parameter $M_{nd}^0$ is set $0.36$, which corresponds to a structure of average density of $0.1$ or $0.9$.

The generated coarse structure is then set as an initial structure for self-supporting topology optimization, together with the chosen printing direction. During the optimization process, in order to prevent the self-supporting constraint hindering the formation of load-carrying structures, the constraint is imposed in a soft way using a similar strategy as previously performed in work~\cite{Guo2014}. Specifically, the value of the tolerance parameter $\epsilon$ in~\eqref{eq-problem} is decreased continuously from a relatively large value to the predefined tolerance $\epsilon_0$ during the course of optimization.

During the process of self-supporting topology optimization, a black-white filter is additionally applied in the last few steps to produce a totally 0-1 density distribution so that the discrete convolution finds exactly the self-supporting elements without the influence of gray elements. This is achieved using the Heaviside projection filter as originally designed by Guest et al~\cite{Guest2004Filter,Andreassen2011Efficient}. Otherwise, gray elements may not be strong enough to support a black element above it, and special care has to be considered~\cite{Lang20161}.

Lastly, in order to have a completely self-supporting structure (without any unsupported elements), the self-supporting constraint is added in a strict way in the last few iteration steps (when the number of unsupported elements is no longer decreasing) via removing non-self supporting elements (whose number is at most 5 is all tests given in this paper). This strategy has an ignorable influence on the final structure's compliance, noticing the nature of optimization approach, i.e. the approach may fluctuate between self-supporting constraints and target optimization and the volume faction constraint. Consequently, all these remaining elements are not essential in determining the structure's physical compliance. This is very different from removing unsupported elements from the support-needed structure in a post-processing step.


\section{Numerical aspects} \label{sec-numerical}
\subsection{Sensitivity analysis}\label{sec-numerical:ss}
The sensitivity of the self-support constraint $U(\brho)$ is straightforward to compute from~\eqref{eq-problem}, and given as:
\begin{equation}\label{eq-sensitivity}
\frac{\partial U}{\partial \rho_e}=
    \begin{cases}
    2\rho_e& \text{if } e \in \M_U \\
    0& \text{if } e \notin \M_U,
    \end{cases}
\end{equation}
where $\M_U$ is the set of unsupported elements. 

Note here that the sensitivity for each element $e$ is only dependent on the density $\rho_e$ of the element itself, without information of density of any other elements, and can be easily implemented in parallel. Previous approaches~\cite{Gaynor2016,Lang20161,Lang20162} have shown skilled techniques in designing self-supporting filters, and the self-supporting sensitivity of each element was expressed in terms of the densities information of the layers below it. The approach has its great freedom in computing self-supporting structure, but also requires a further improvement to overcome it inhabitation of parallel processing, as also explained by the authors~\cite{Gaynor2016,Lang20161,Lang20162}.

Derivations of the sensitivities of the objective function $\c(\brho)$ or of the volume constraint $V(\brho)$, involved in~\eqref{eq-problem}, are totally the same as those done in previous studies~\cite{Sigmund2001A,Sigmund2013}. Integrating these sensitivities with thickness control can be achieved using the Heaviside filter~\cite{Guest2004Filter}, as will be demonstrated in Section~\ref{sec-example}. Details are not further explained here.

\subsection{Discrete convolution for efficient unsupported element detection}\label{sec-numerical:conv}
Computing the sensitivity~\eqref{eq-sensitivity} of the self-supporting constraints requires detecting the set $\M_U$ of all the unsupported elements. They can be easily detected via enumerating all the discrete elements $e$ of $\M$ not satisfying the property in~\eqref{eq-ss2D} or~\eqref{eq-ss3D}. Such process is however very time-consuming. In order to further accelerate this process, a novel convolution operator is designed for such detections, which acceleration the process of detecting the unsupported element with a speedup of 100 times as compared with the approach of directly enumerating them element-wise. 

pGiven a discrete structure $\M$ of size $N\times M$ in 2D, we can see from~\eqref{eq-ss2D} that an element $e(n,m)\in\M$ is supported if the summation of the densities of its supporting elements is larger than zero, or specifically,
\[
\sum_{n-1\leq r\leq n+1} \rho(r,m-1)> 0.
\]

\begin{figure}[tb]
\centering
\includegraphics[width=0.1\textwidth]{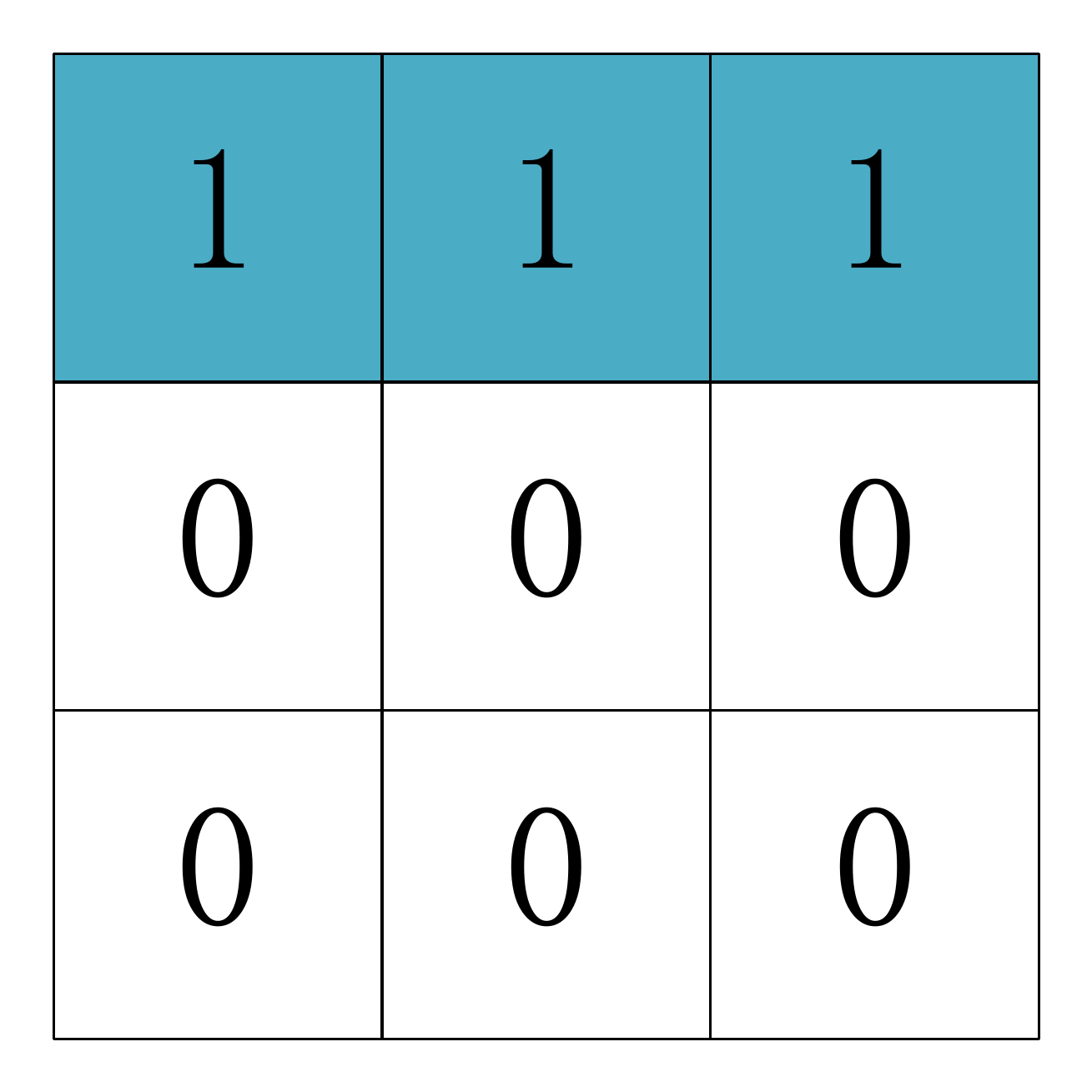}
\caption{The 2D self-supporting convolution kernel matrix $\H$ to detect supported elements for an overhang  angle of $45^{\circ}$.}
\label{fig-2dconv}
\end{figure}
\begin{figure}[bt]
\centering
\includegraphics[width=0.35\textwidth]{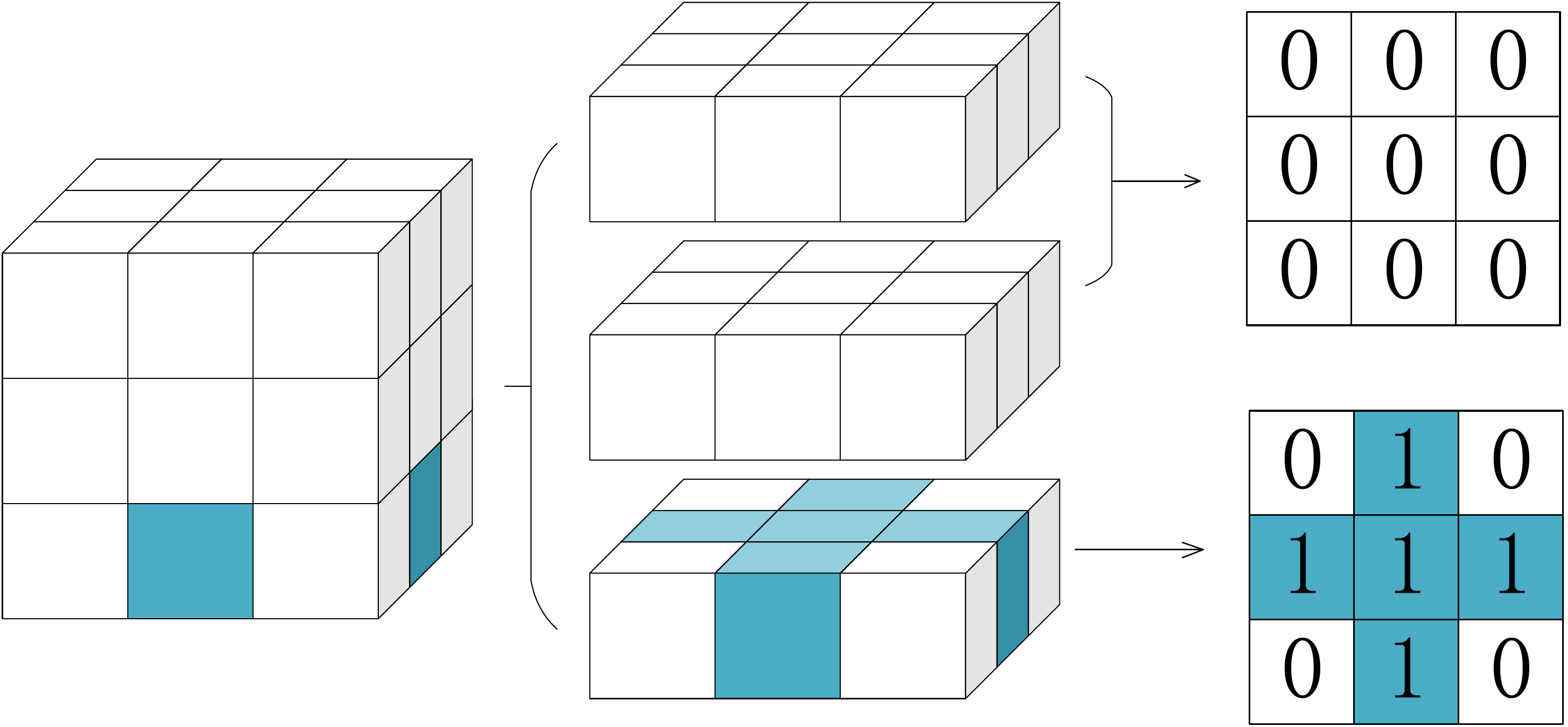}
\caption{The 3D self-supporting convolution kernel matrix $\H$ to detect supported elements for an overhang  angle of $45^{\circ}$.}
\label{fig-3dconv}
\end{figure}

\begin{figure*}[bt]
\centering
\includegraphics[width=0.6\textwidth]{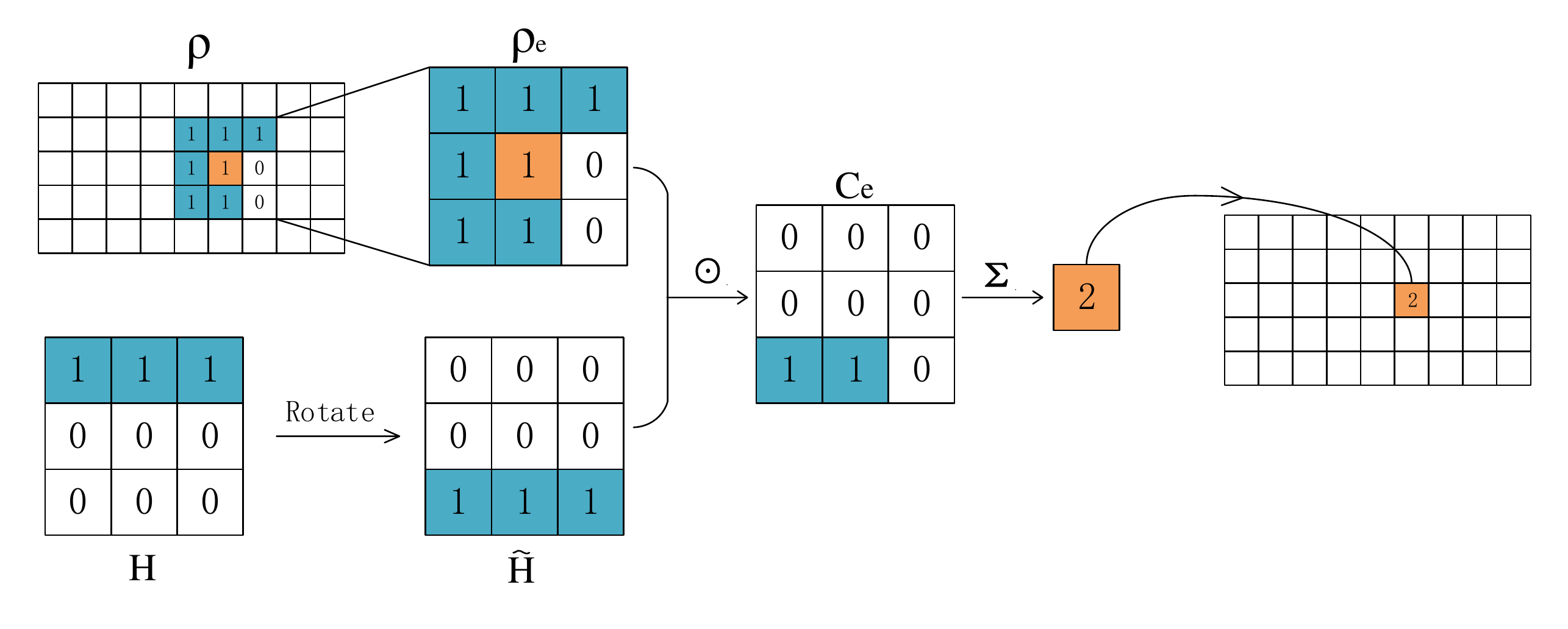}
\caption{The discrete convolution procedure in 2D for detecting supported elements using the designed self-supporting Kernel matrix $\H$ in Fig.~\ref{fig-2dconv}.}
\label{fig-2dconv-comp}
\end{figure*}

The newly introduced \emph{self-supporting convolution operator} is designed based on this observation. Specifically, suppose the overhang angle is $45^{\circ}$. The associated 2D self-supporting kernel matrix $\H$ of size $3 \times 3$ is defined in Fig.~\ref{fig-2dconv}.
A new matrix $\brho_S$ is then computed via performing the convolution between the density matrix $\brho$ and $\H$, or
\begin{equation}\label{eq-sp-filter}
\tilde{\brho}_S = \text{sign}(\brho* \H),
\end{equation}
where $\brho*\H$ is the convolution between matrices $\brho$ and $\H$, whose $(n,m)$ element is defined as
\[
(\brho*\H)(n,m) = \sum_{i=-1}^{1} \sum_{j=-1}^{1} \rho(n-i,m-j) \cdot \H(i+1,j+1),
\]
and the sign function
\begin{equation}\label{eq-sign}
\text{sign}(x)=
\begin{cases}
0& \text{if  } x = 0,\\
1& \text{if  } x > 0.
\end{cases}
\end{equation}
Correspondingly, we have the set of supported elements $\M_S$ of the discrete structure $\M$,
\begin{equation}\label{eq-sse}
\M_S=\{ e(n,m)\ |\ \tilde{\brho}_S(n,m)=1\},
\end{equation}
for $\tilde{\brho}_S$ defined in~\eqref{eq-sp-filter}. 

The above expression assumes a 0-1 density distribution of $\brho$, while in the SIMP approach as studied here, the density matrix $\brho$ usually has entry value ranging from 0 to 1. A specific value usually needs to be set to replace 0 in the sign function~\eqref{eq-sign} for practical applications.

The basic procedure of the convolution computation is further shown in Fig.~\ref{fig-2dconv-comp} and explained below. For each element $e$ under consideration, a $3\times3$ matrix $\brho_e$ centering at $e$ is selected.
This is then followed by its Hadamard product, i.e. the element by element product between the matrices,
\begin{equation}\label{eq-hp}
C_e=\tilde{\H}\odot \brho_e,
\end{equation}
where $\tilde{\H}$ is the rotation of matrix $\H$ at a degree of $180^{\circ}$. The convolution value of element $e$ is the summation of all the values in the derived matrix $C_e$.

The above procedure works for every element $e$. For the boundary elements, an additional loop of void elements are added. Note also that the bottom elements are always taken as supported considering the fact that they are always supported by the baseboard of the fabrication device.

Once the set $\M_S$ of supported elements of $\M$ is determined from~\eqref{eq-sse}, the set of unsupported element is derived consequently,
\begin{equation}
\M_U = \M \setminus \M_S.
\end{equation}

The convolution procedure in 3D is similar to that in 2D, and the corresponding kernel matrix $\H$ is shown in Fig.~\ref{fig-3dconv}. Extension of the approach to general overhang angle will be laterp explained in~\ref{sec-numerical:angle}. 

We also compare in Table~\ref{tab:conv} the computational time in detecting the supported elements using the proposed self-supporting convolution operator and using direct enumeration element by element. Almost a 2 order of speedup is observed from the results. The time of detecting supported elements is not ignorable compared with FE computations. For example, one step of FE for the size of $600\times 400$ just takes 0.4163 seconds, and the enumeration time takes 0.5214 seconds. This is very important for the practical usage of the proposed approach, particularly on 3D complex structures with millions of elements, considering that FE computations can be implemented in parallel and accelerating the detection of supported elements then becomes important.  

\begin{table}[H]
\centering\small
\begin{threeparttable}
\caption{\label{tab:conv} Time-cost comparison between direct enumeration and using convolution for detecting supported elements in 2D and 3D cases}
\begin{tabular}{lcccc}
\toprule
Domain size & Enumeration (s) & Convolution (s) & Speedup \\
\midrule
$80 \times 40$ & 0.0094 & 0.0001 & 94 \\
$320 \times 80$ & 0.0380 & 0.00035 & 106.16 \\
$600 \times 400$ & 0.5214 & 0.0034 & 152.11 \\
$40 \times 40 \times 20$ & 0.1419 & 0.0010 & 139.13 \\
$100 \times 100 \times 100$ & 4.7974 & 0.0324 & 147.82 \\
$500 \times 200 \times 100$ & 114.7641 & 0.7605 & 150.88 \\
\bottomrule
\end{tabular}
\end{threeparttable}
\end{table}

Note that various convolution operators have been designed and used as filters in topology optimization for design control, for example for removing checkerboard patterns or thickness control~\cite{D1995Checkerboard}. They usually aim to compute an element's density or sensitivity via averaging those of the elements around it. In contrast to these researches, the convolution is used here for detecting unsupported elements. It does not change the element density. 

\begin{figure}[tb]
\centering
\subfigure[Supporting elements]{\includegraphics[width=0.23\textwidth]{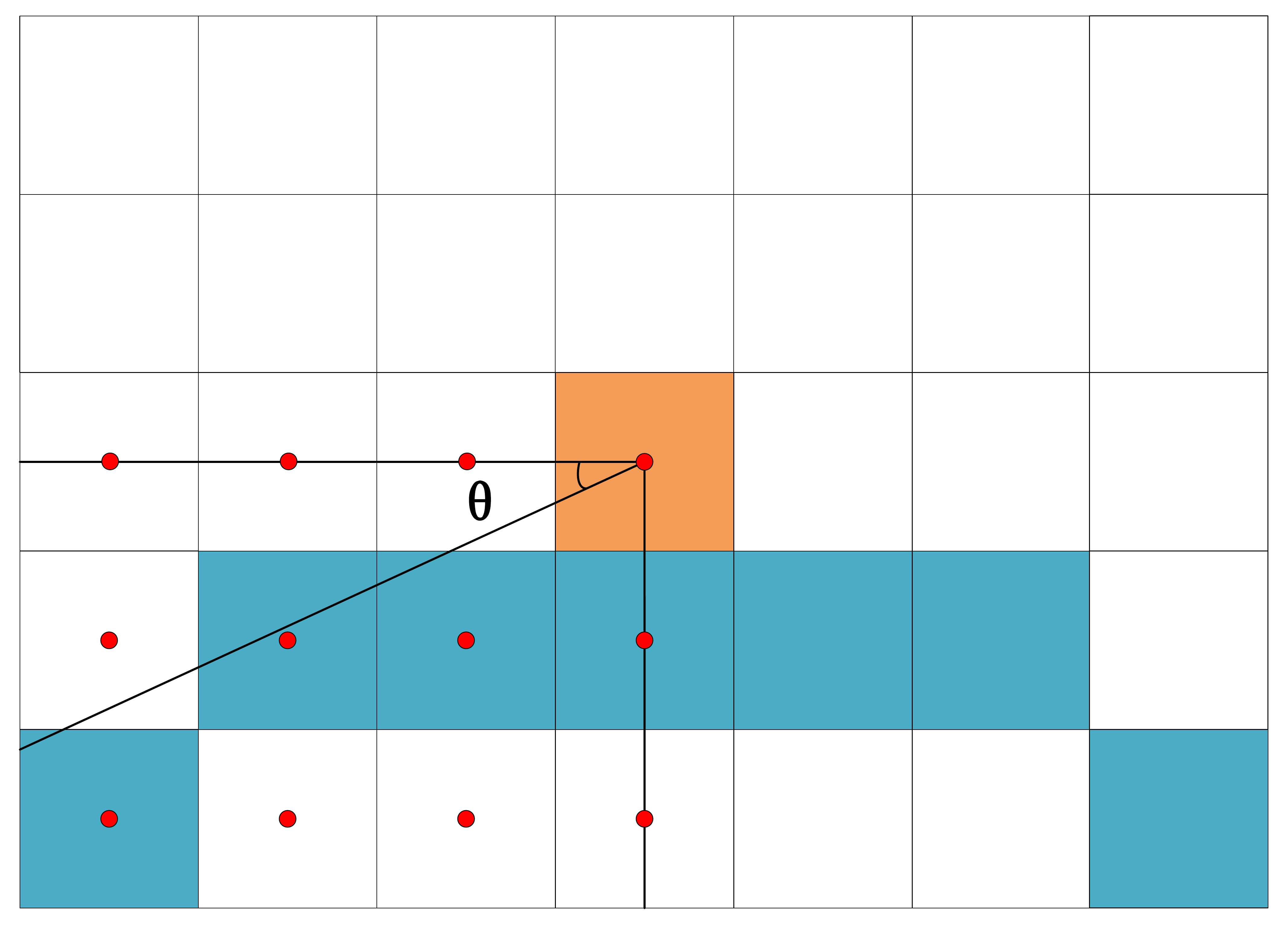}}\quad
\subfigure[Convolution Kernel matrix $\H$]{\includegraphics[width=0.23\textwidth]{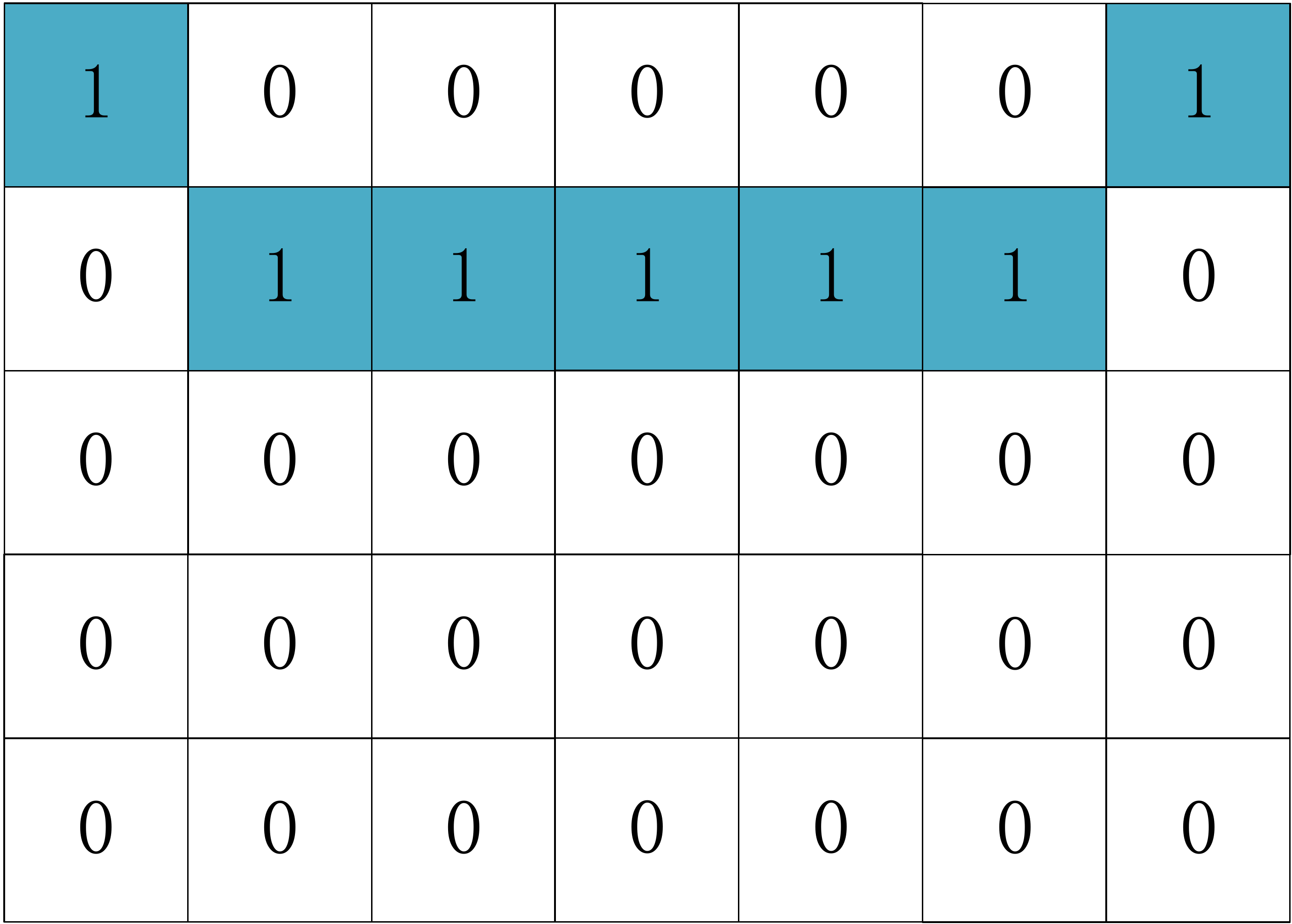}}
\caption{Procedure for building a self-supporting convolution Kernel matrix for a general overhang angle $\theta$.}
\label{fig:irr-theta}
\end{figure}

\subsection{Extension to general overhang angle}\label{sec-numerical:angle}
The above described procedure is mainly described for an overhang angle of $45^{\circ}$ for ease of explanations. Its extension to a general overhang angle $\theta$ is further explained below. The only difference is the construction of the kernel matrix $\H$ involved in~\eqref{eq-hp}.

As shown in Fig.~\ref{fig:irr-theta}(a), given an overhang angle $\theta$, a straight line $L$ passing through the center of an element $e$ and with a slope angle $\theta$ is drawn. Then the first element in each column whose centers are below line $L$ are taken as the supporting elements with respect to element $e$. Their densities are set $1$ and others' $0$,  which together determines a matrix $\bar{\H}$. Rotating $\bar{\H}$ with respect to the center $e$ at a degree of $180^{\circ}$ gives the convolution kernel matrix $\H$,  as shown in Fig.~\ref{fig:irr-theta}(b). The above procedure works for building convolution Kernel using multiple layers; the more layers taken the more accurate of the built kernel matrix for detecting the supported elements.

\section{Examples}\label{sec-example}


Extensive 2D and 3D examples are performed to test performance of the proposed approach. For illustration purposes, the material, load and geometry data are chosen to be dimensionless. The Young's modulus and Poisson's ratio of the solid material are set as $E = 1$ and $\upsilon = 0.3$ for all examples. The penalization factor is set to a value of $3$. The minimal thickness is set to be 2, and the overhang angle is set to $45^{\circ}$ if not explicitly specified. The printing direction is selected during each optimization process and marked in the example figure. The 2D examples were implemented in Matlab, and the 3D examples were implemented in C++ and GPU for parallel computations on a computer
of 3.2G CPU, 8.0G RAM and GeForce GTX 970 Graphic card.

The examples include the classical Cantilever beam, MBB, a 2D square example  and three 3D examples. The Cantilever beam is used to illustrate various aspects of the approach: basic performance on a rectangular domain or a general domain, iteration process, thickness control, different volume fractions, different overhang angles, different types of external forces. The MBB shows the approach's ability in handling constraints of multiple print directions. The 2D square example demonstrates the approach's performance in case of complex topological structure obtained at distributed external forces. The 3D examples are further used to demonstrate the approach's ability in handling complex 3D models of millions of DOFs via parallel implementation.

Following previous studies~\cite{Gaynor2014,Gaynor2016,Lang20161,Lang20162}, we measure the ability of a self-supporting topology optimization approach in maintaining the structure's physical performance using the compliance ratio
\begin{equation}\label{eq-ratio}
\frac{C}{C_{ref}},
\end{equation}
where $C_{ref},C$ is respectively the compliance of the structure computed with or without considering self-supporting constraint.

The computational results for 2D examples are first summarized in Table~\ref{table:comp}; cases of 3D examples are explained later.
\begin{table*}[tb]
\centering\small
\begin{tabular}{lccccccccc}
\toprule
&$\#\M$ & $\#\M_U$  & $C_{ref}$  & $C$ & $C/C_{ref}$  \\
\midrule
Beam  & 9000 & 24 & 92.7 & 92.8 & 100.11\% \\
Beam (hole) & 7755 & 25 & 115.4 & 117.4 & 101.73\%\\
Beam ($r=1.5$) & 9000 & 24        & 92.7      & 92.8 & 100.11\%\\
Beam ($r=2$)   & 9000 & 18        & 92.4      & 92.8 & 101.43\%\\
Beam ($r=3$)   & 9000 & 12        & 92.6      & 92.9 & 100.32\%\\
Beam (concentrated) &14400 & 21   & 322.9     & 323.0 & 100.31\%\\
Beam (distributed)  &14400 & 12   & 255.4     & 256.1 & 100.27\%\\
Beam (mixed)        &14400 & 60   & 31971.1   & 32514.3 & 101.70\%\\
Beam (vf=0.6)       & 9000 & 24 & 92.7  & 92.8  & 100.11\% \\
Beam (vf=0.5)       & 9000 & 97 & 105.8 & 106.8 & 100.95\%\\
Beam (vf=0.4)       & 9000 & 37 & 127.4 & 128.3 & 100.71\%\\
Beam (vf=0.25)      & 9000 & 55 & 196.8 & 202.9 & 103.10\%\\
Beam (Angle=30)     & 9000 & 12  & 105.8 & 113.6 & 107.37\%\\
Beam (Angle=45)     & 9000 & 97 & 105.8 & 106.8 & 100.95\% \\
Beam (Angle=60)     & 9000 & 425& 105.8 & 141.3 & 133.55\%\\
MBB                 & 38400 & 1874        & 185.7     & 191.3 & 103.02\%\\
Square              &22500 & 474  & 1312.8    & 1544.7 & 117.66\%\\
\bottomrule
\end{tabular}
\caption{\label{table:comp} Summary of the numerical results for various tested 2D examples: $\#\M$, element number of the design domain; $\#\M_U, C_{ref}$, number of unsupported elements of structured computed from topology optimization without considering self-supporting constraints and its associated compliance; $C$, compliance of the self-supporting structure computed using the proposed approach. The (Cantilever) beam examples are described in Figs.~\ref{fig:longbeam} and~\ref{fig:forces}, the MBB example in Fig.~\ref{fig:mbb}, the 2D squares in Fig.~\ref{fig:square}, the 3D examples in Figs.~\ref{fig:wheel},\ref{fig:cantilever} and~\ref{fig:desk}}.
\end{table*}

\subsection{Cantilever beam example}
\begin{figure}[tb]
\centering
\subfigure[Rectangular domain]{\includegraphics[width=0.35\textwidth]{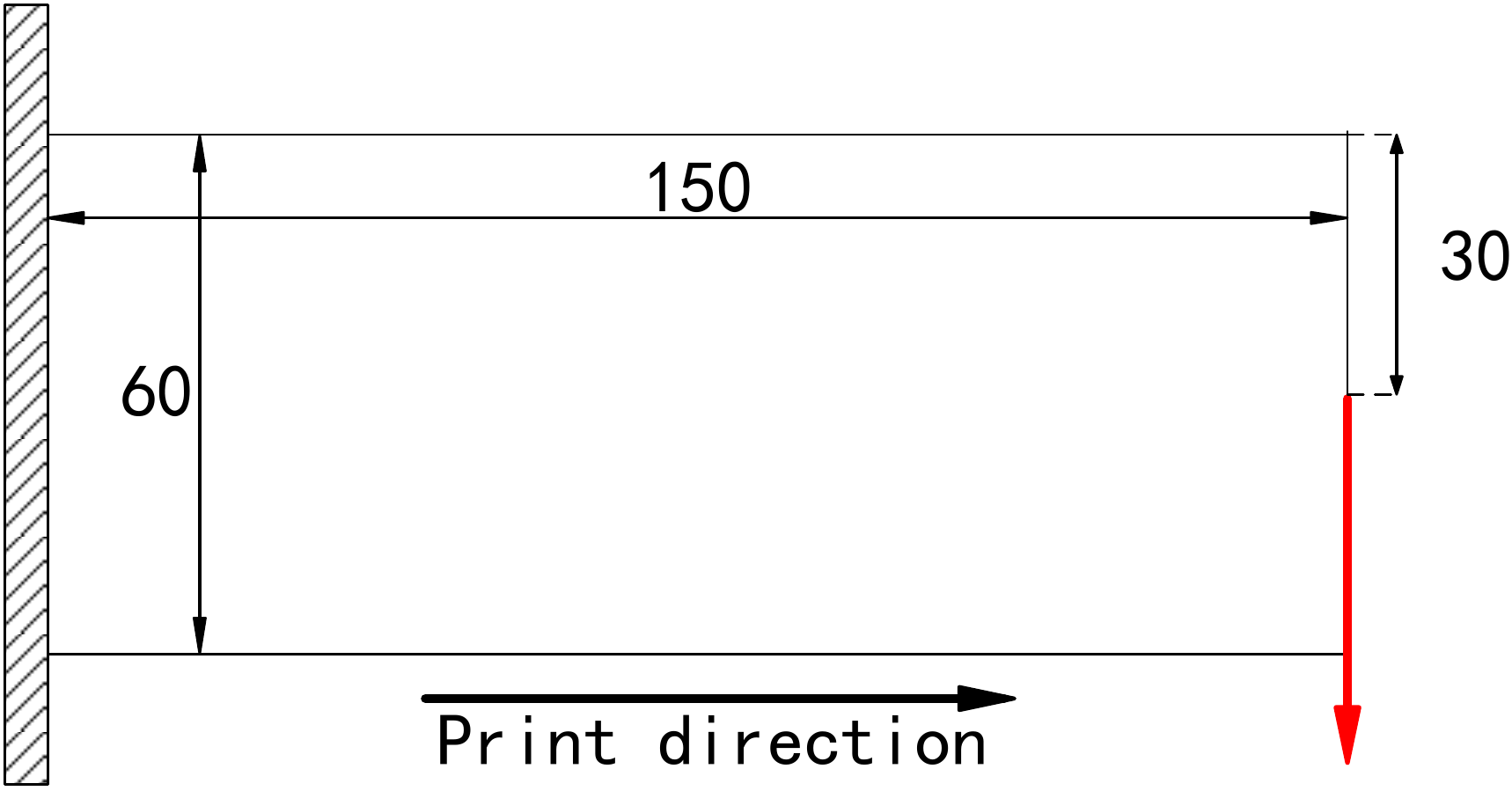}}
\subfigure[General domain]{\includegraphics[width=0.35\textwidth]{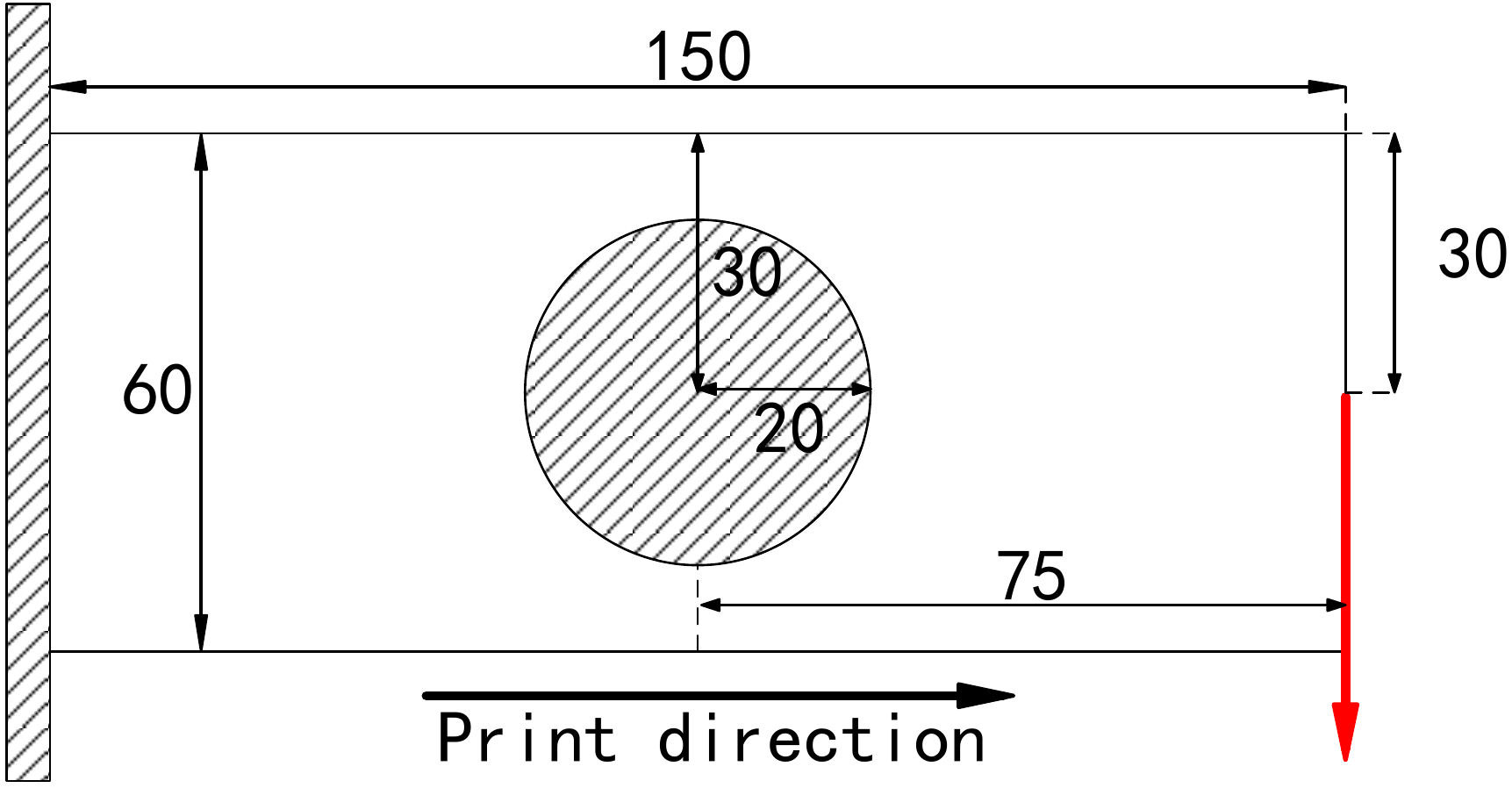}}
\caption{The Cantilever beam examples with a minimum thickness of $1.5$.}
\label{fig:longbeam}
\end{figure}

The Cantilever beam, as shown in Fig.~\ref{fig:longbeam}, is first tested. The model on the left has a $150 \times 60$ rectangular domain, and has a target volume fraction of $0.6$. The model on the right has general domain made via cutting a circular hole within the left one, and has a target volume fraction of $0.5$. Both models are fixed on the left edge with an external force exerted on the middle point of its right edge. The print direction is determined from left to right.

\begin{figure}[tb]
\centering
\subfigure[Support-needed]{\includegraphics[angle=90,width=0.1\textwidth]{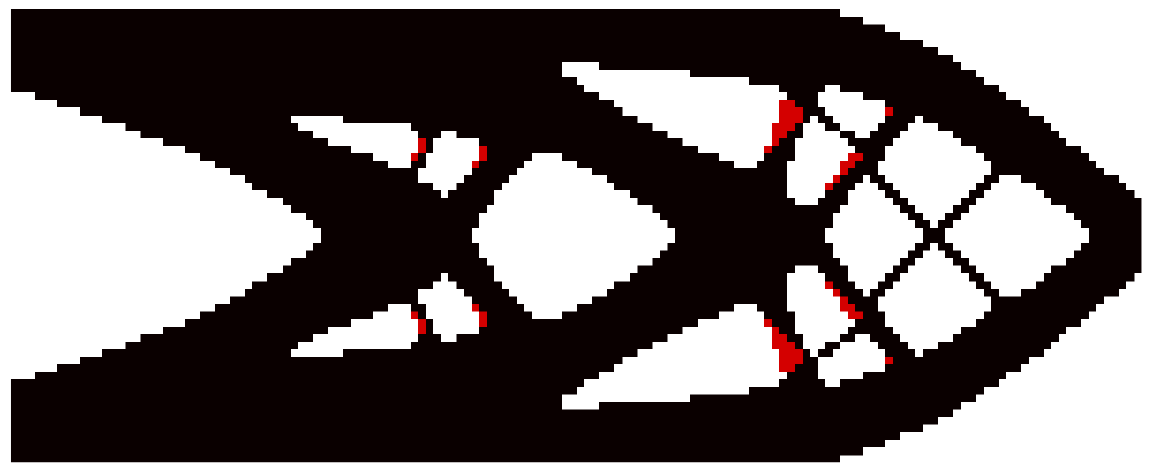}}\quad
\subfigure[Self-supporting]{\includegraphics[angle=90,width=0.1\textwidth]{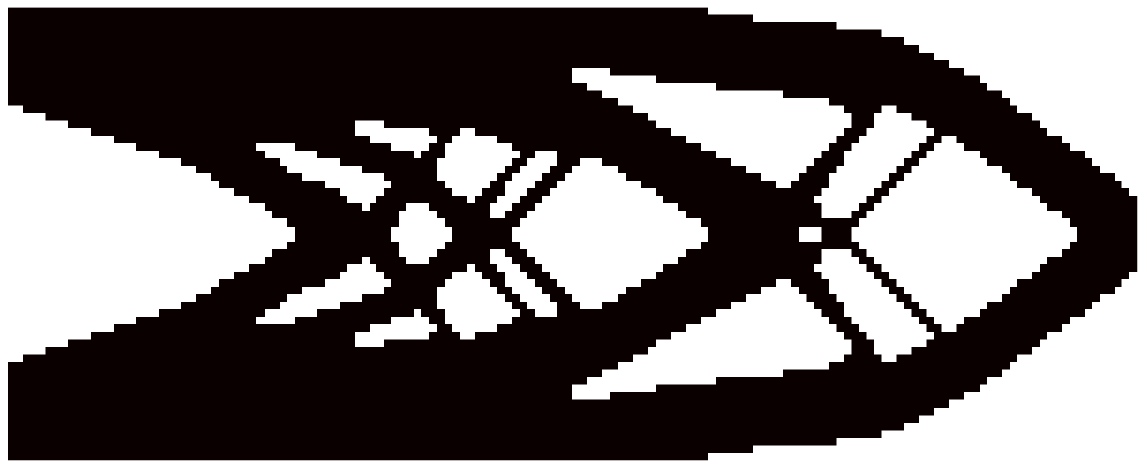}}\quad
\subfigure[Support-needed]{\includegraphics[angle=90,width=0.1\textwidth]{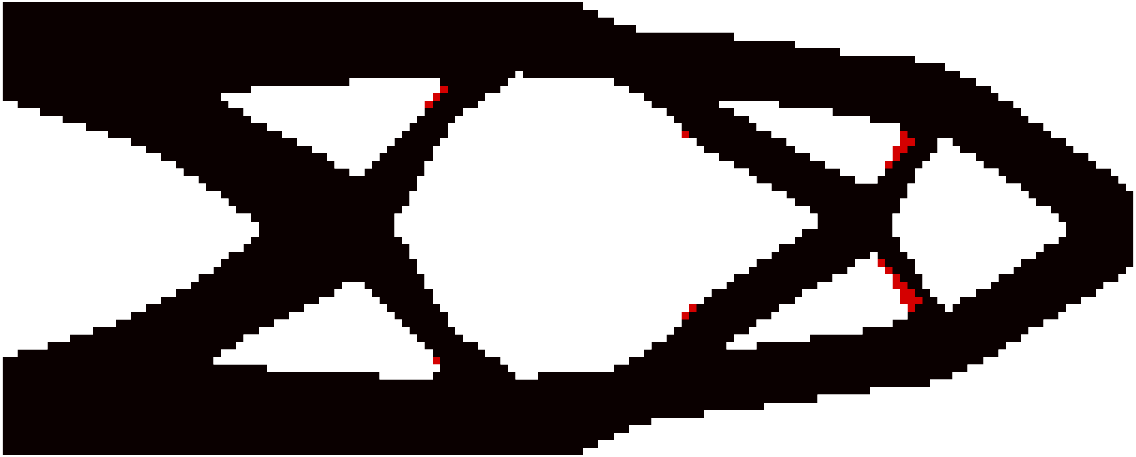}}\quad
\subfigure[Self-supporting]{\includegraphics[angle=90,width=0.1\textwidth]{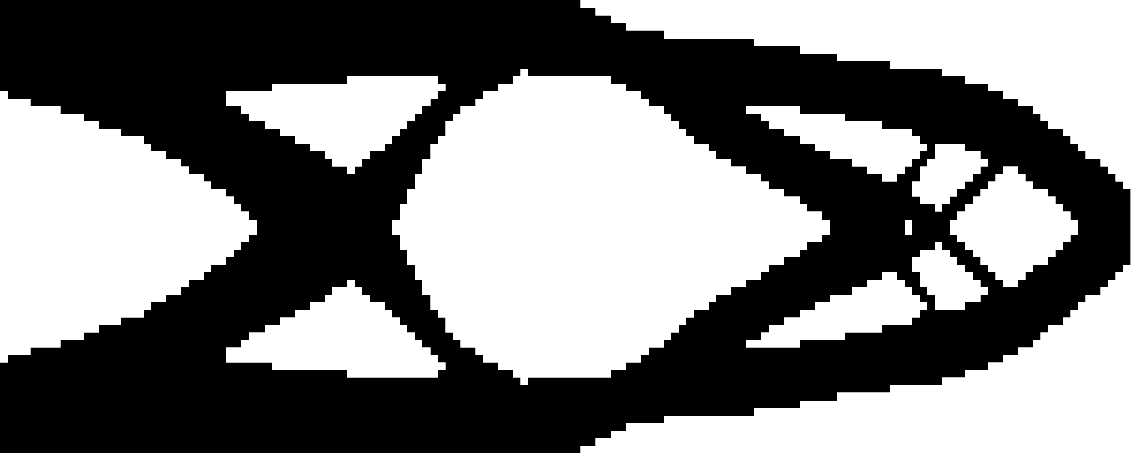}}
\caption{Computational results for the Cantilever beam example in Fig.~\ref{fig:longbeam}. See also Table~\ref{table:comp} for more details.}
\label{fig:longbeam:basics}
\end{figure}

Without considering the self-supporting structure, the structure in Fig.~\ref{fig:longbeam:basics}(a),(c) are obtained where the elements in red are those that cannot be successfully printed out. The proposed self-supporting topology optimization approach results in the structures in (b) and (d), both of which do not contain any unsupported elements. We can see from the results that the range containing unsupported elements in (a),(c) moves upward in (b),(d) to adapt the requirement of self-supporting. In addition, it is also very interesting to notice that various parts of (b) or (d) are different from those of (a) or (c), for the structures' maximal physical performance, although simultaneously maintaining their overall structures. The structures computed with or without self-supporting constraints have a very close compliance, of a compliance ratio respectively of $100.11\%$ and $101.73\%$ for the left and right examples. 

In handling the right model of a general design domain, we work on the rectangular domain following the procedure below. In each step of the optimization iteration, the density of each element within the circular domain is set $0$, and then the convolution operation (detailed in Section~\ref{sec-numerical:conv}) is performed in the whole rectangular domain to detect the unsupported elements. The above two steps are repeated until convergence.


\subsubsection{Iteration performance}
\begin{figure}[tb]
\centering
\subfigure[step 20]{\includegraphics[angle=90,width=0.142\textwidth]{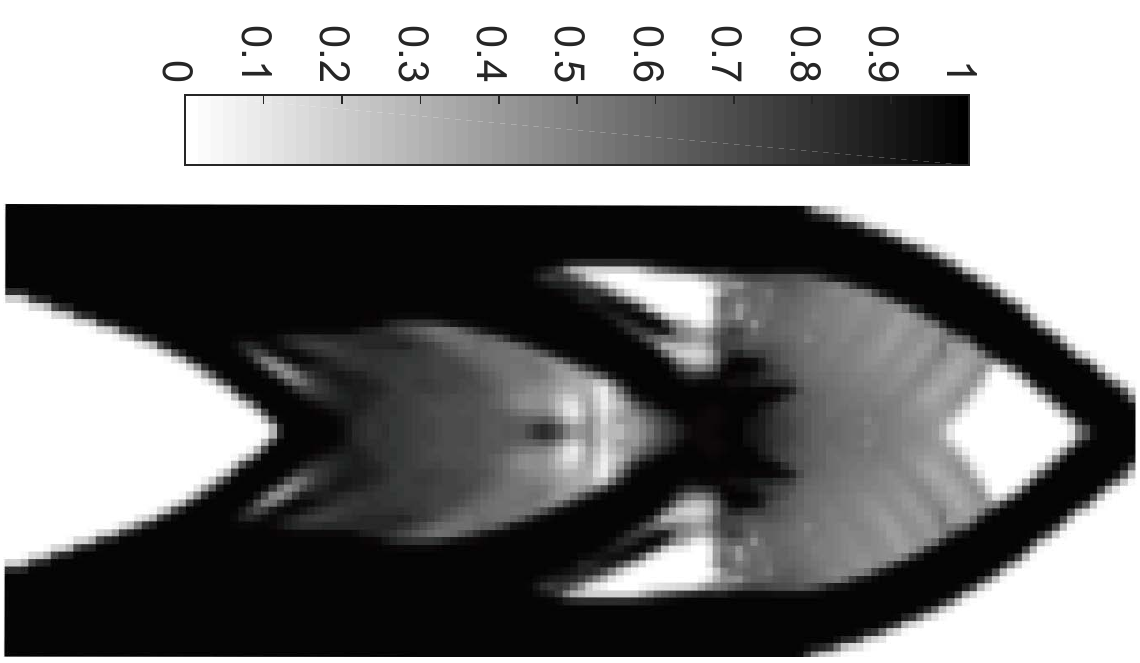}}\quad
\subfigure[step 44]{\includegraphics[angle=90,width=0.1\textwidth]{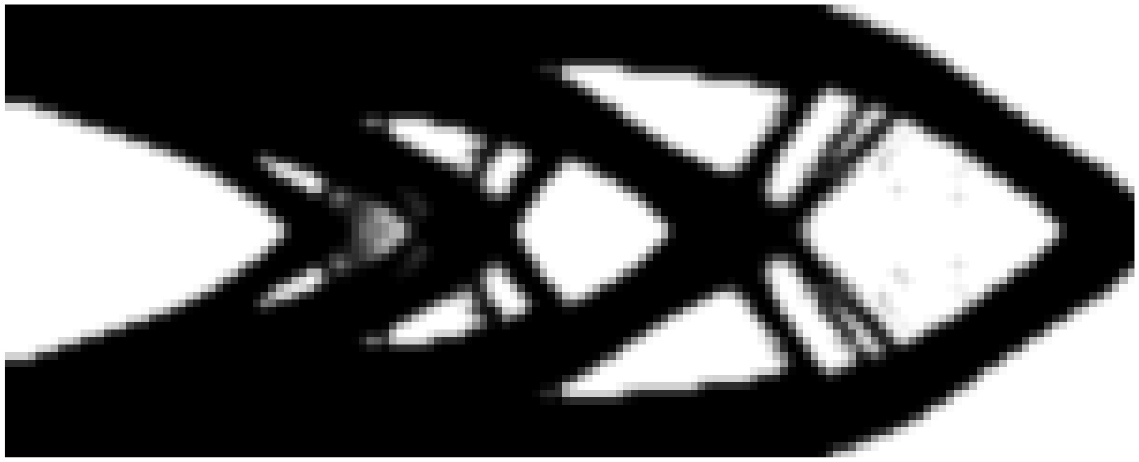}}\quad
\subfigure[step 45 (22)]{\includegraphics[angle=90,width=0.1\textwidth]{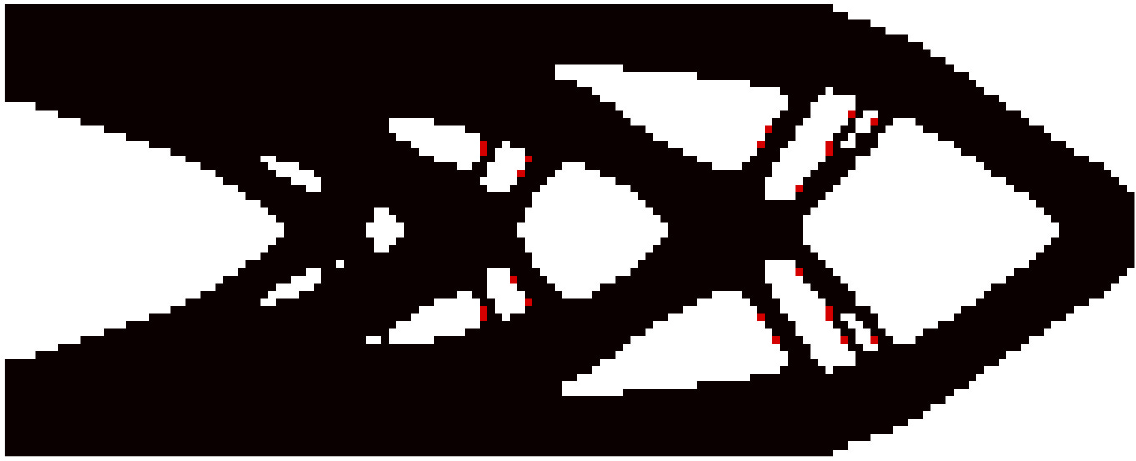}}\\
\subfigure[step 54 (12)]{\includegraphics[angle=90,width=0.1\textwidth]{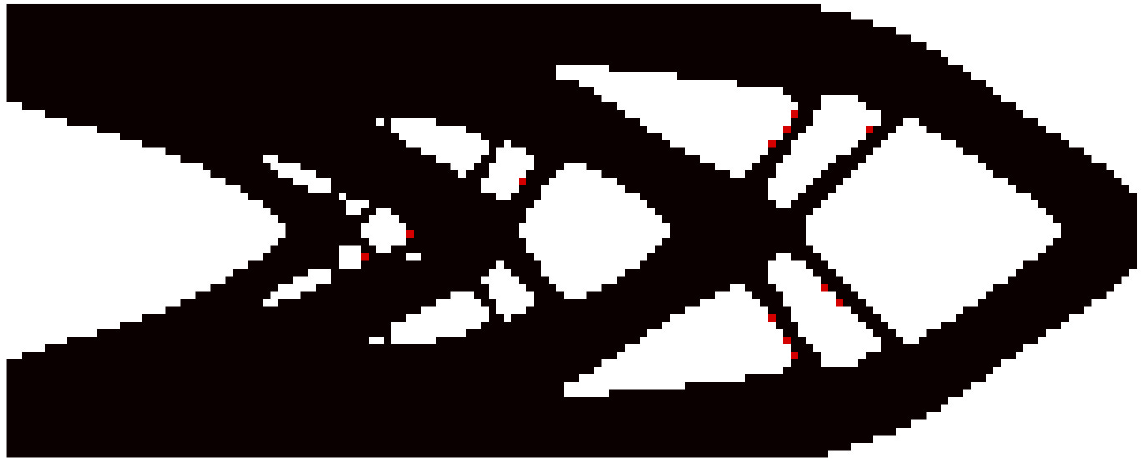}}\quad
\subfigure[step 82 (6)]{\includegraphics[angle=90,width=0.1\textwidth]{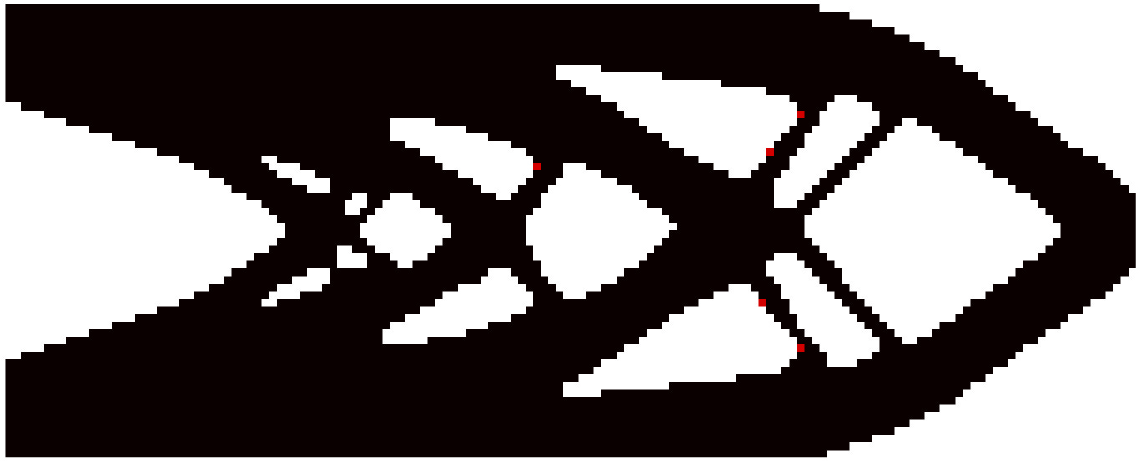}}\quad
\subfigure[step 101 (3)]{\includegraphics[angle=90,width=0.1\textwidth]{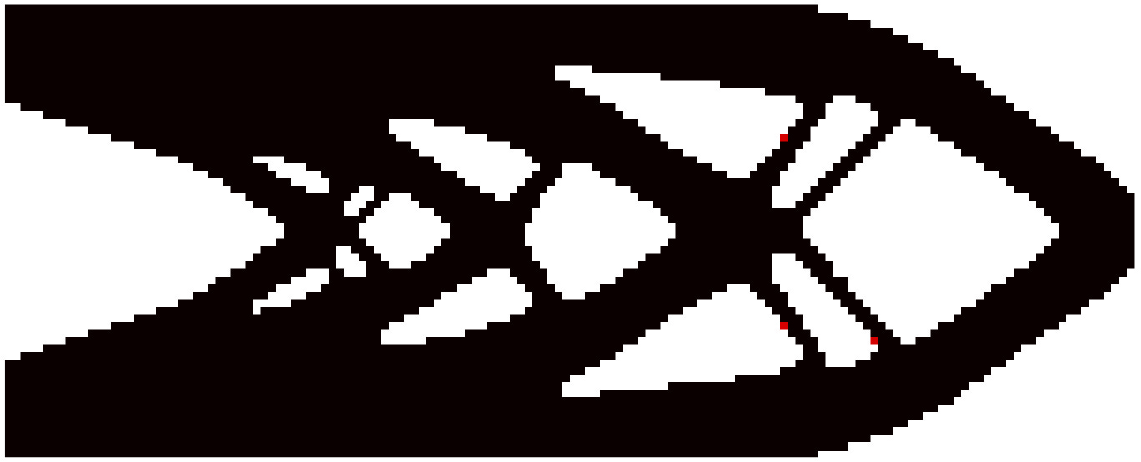}}\quad
\subfigure[step 166 (0)]{\includegraphics[angle=90,width=0.1\textwidth]{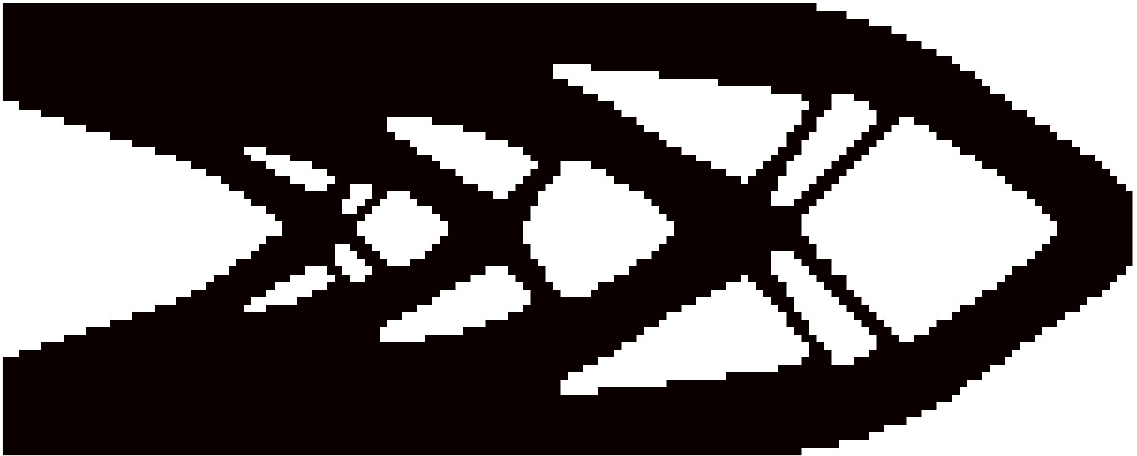}}
\caption{Key frame figures at different iteration steps, where the number in brackets stands for the number of unsupported elements.}
\label{fig:int:top}
\end{figure}

\begin{figure*}[tb]
\centering
\subfigure[Unsupported elements]{\includegraphics[width=0.35\textwidth]{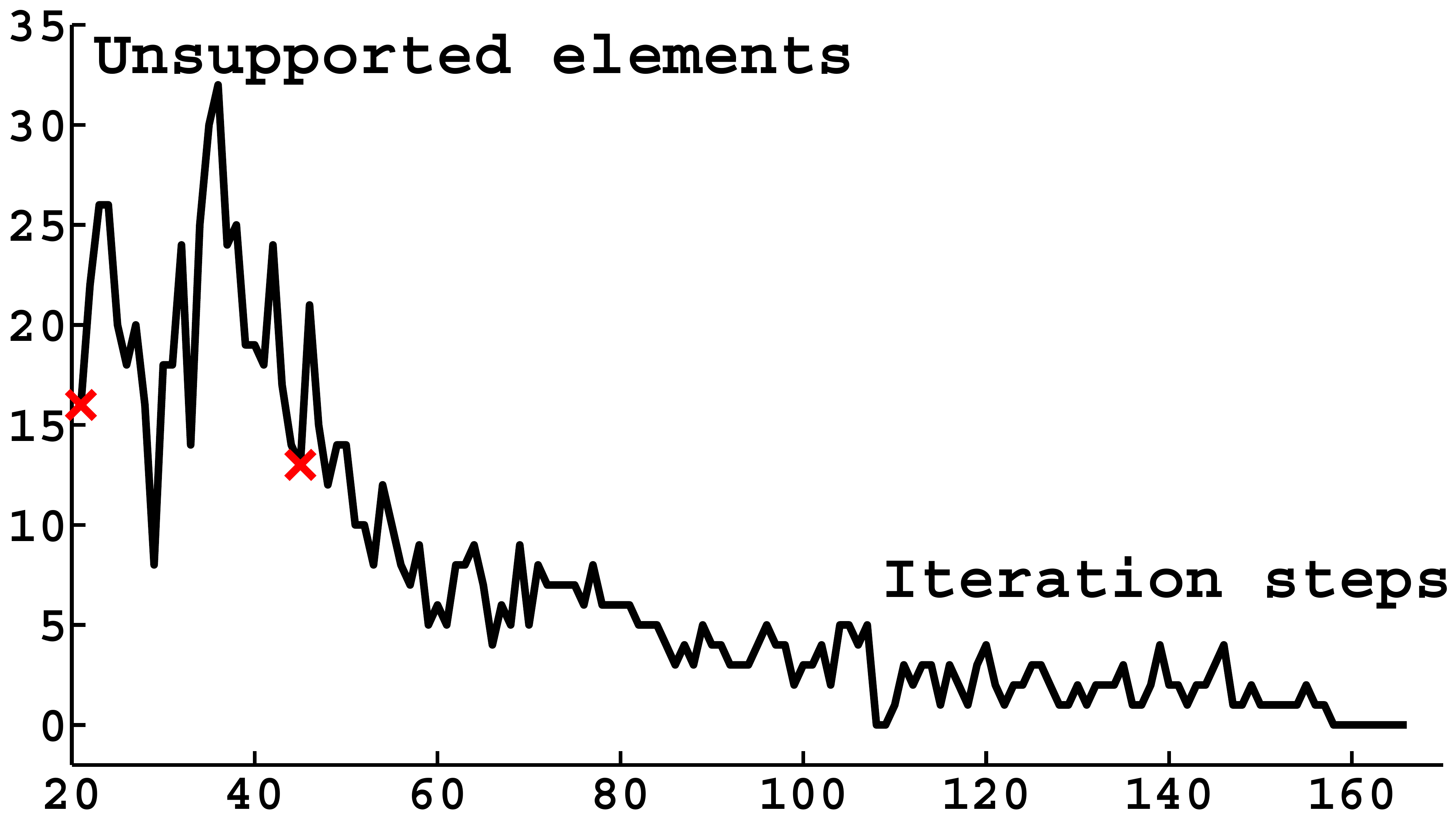}}
\subfigure[Compliance]{\includegraphics[width=0.3\textwidth]{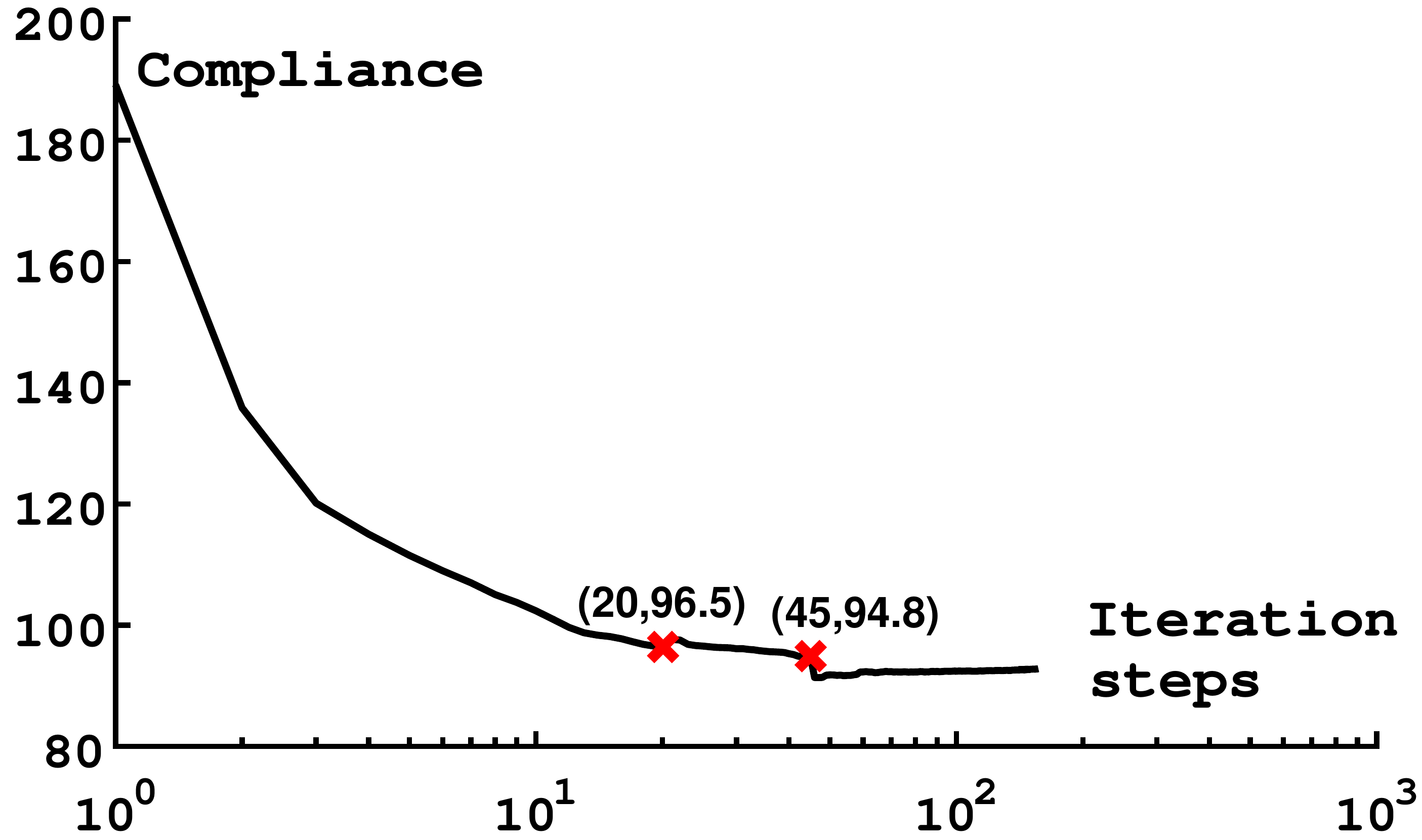}}  
\subfigure[Volume fraction]{\includegraphics[width=0.3\textwidth]{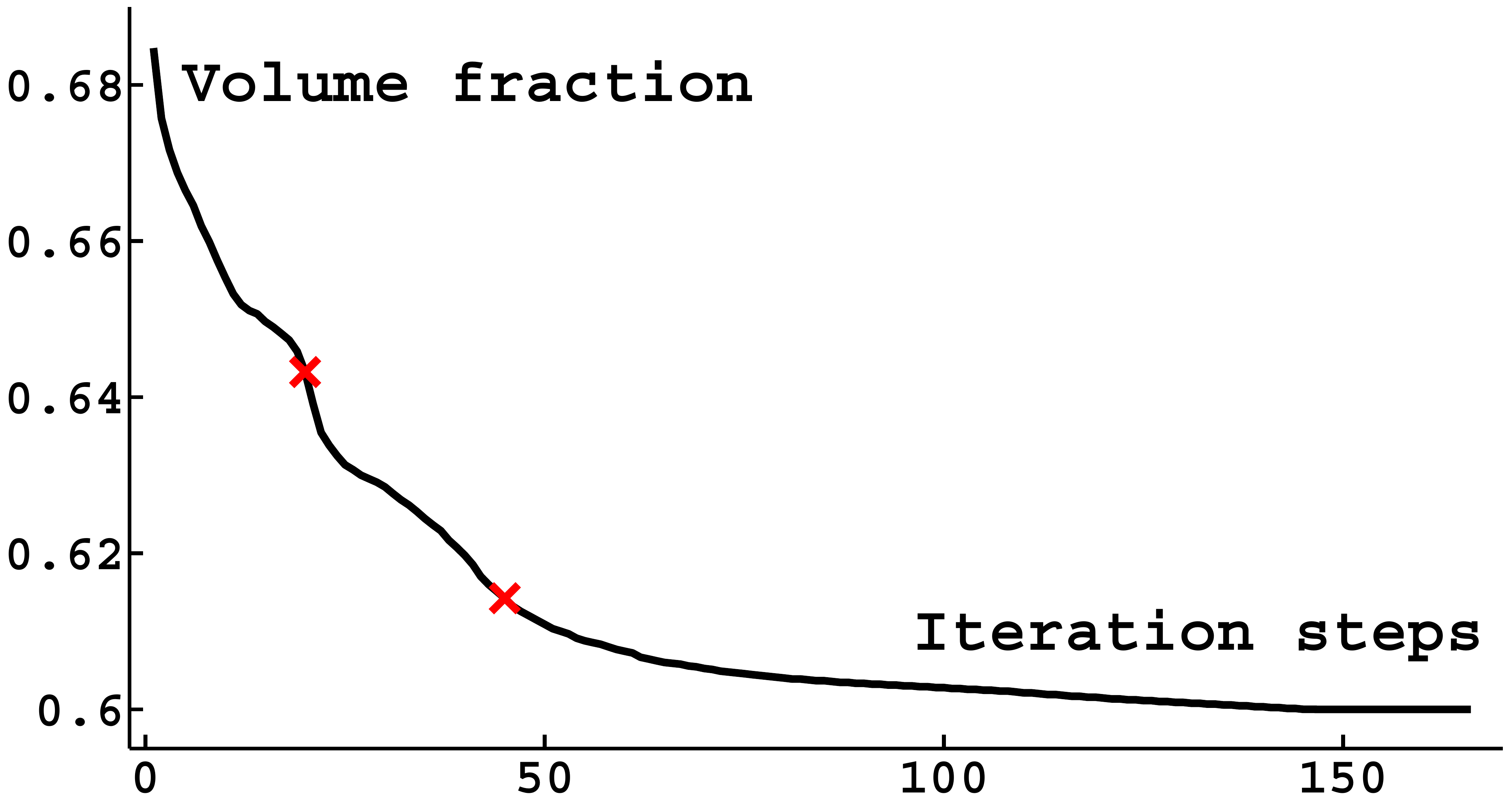}}
\caption{Variations of the structure's number of unsupported elements, compliance and volume fraction during the optimization iteration steps. }
\label{fig:int:ecc}
\end{figure*}

The iteration process of the example given in Fig.~\ref{fig:longbeam}(a) is further explored by examining the variations of the structure's topology, the number of unsupported elements and the compliance and volume fraction of the derived structure, as shown in Figs.~\ref{fig:int:top} and~\ref{fig:int:ecc}.

The iteration process is divided into the following main steps (see also Fig.~\ref{fig:int:top}). Firstly, a topology optimization step without considering the self-supporting constraint is performed, and results in the "gray" structure in (a). After this, the relaxed self-supporting constraint is added in the optimization iteration step, producing a structure in (b). The derived "gray" structure is then transformed into a black-white structure using the Heaviside project filter, as given in (c). After this, the self-supporting topology optimization process is iterated to reduce the number of unsupported elements while simultaneously optimizing its physical performance and maintaining its volume fraction, producing the structures in (d),(e),(f), and ultimately the final structure in (g). The unsupported elements are marked red in Figs.~\ref{fig:int:top}(c)-(g), and illustrated in the caption. Their number is gradually decreased during the optimization iteration process.

Figure~\ref{fig:int:ecc} shows the variation of the structure's number of unsupported elements, target compliance and volume fraction in (a),(b) and (c) respectively. As can be seen, as the iteration step increases, the number of unsupported elements decrease until finally reaching zero. However, fluctuations in the number of unsupported elements may happen during the iteration process. The structures' compliance and volume fraction decrease and finally reaches to a stable state.

\subsubsection{Thickness control}
\begin{figure}[tb]
\centering
\subfigure[$r_{min}=1.5$]{\includegraphics[angle=90,width=0.1\textwidth]{long-beam-step-166.pdf}}\quad
\subfigure[$r_{min}=2$]{\includegraphics[angle=90,width=0.1\textwidth]{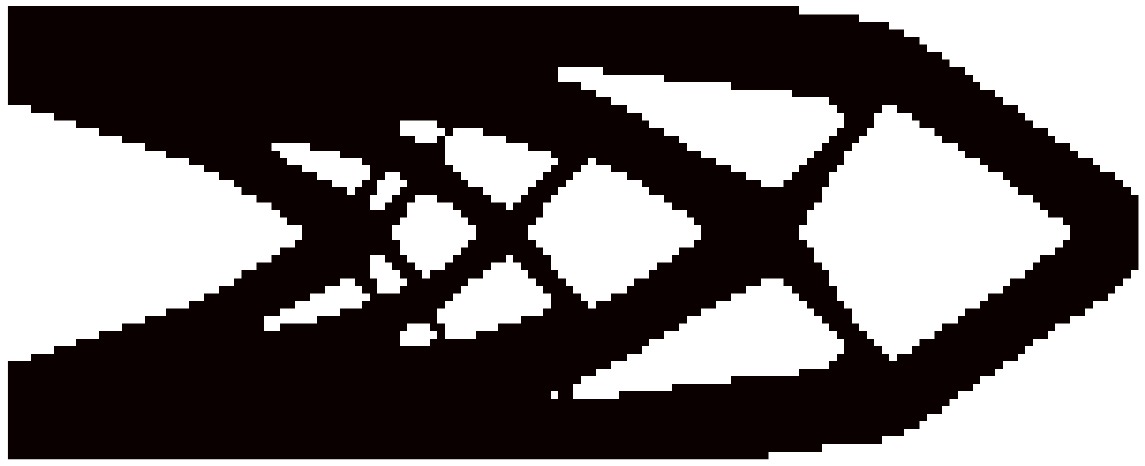}}\quad
\subfigure[$r_{min}=3$]{\includegraphics[angle=90,width=0.1\textwidth]{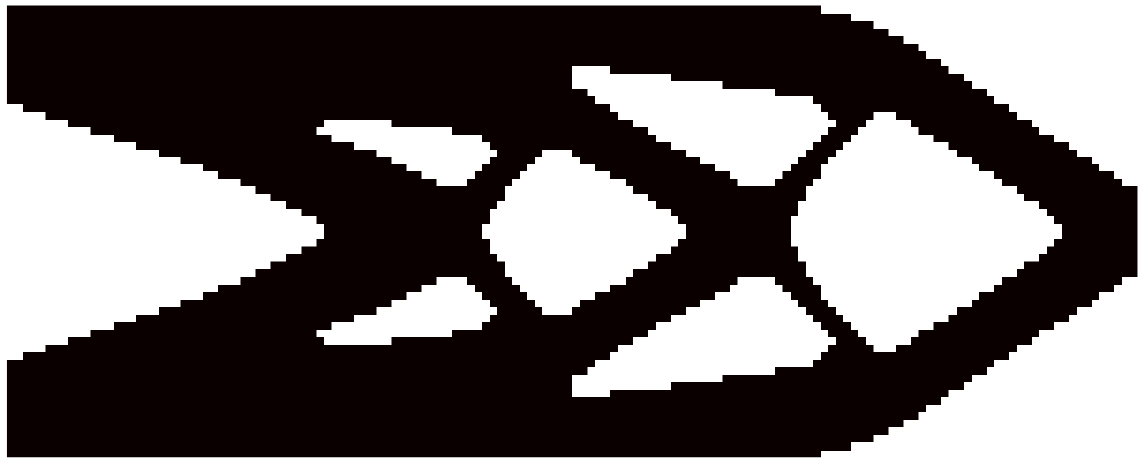}}
\caption{Numerical results for the Cantilever beam example in Fig.~\ref{fig:longbeam}(a) at different thicknesses of $1.5,2$ and $3$. See also Table~\ref{table:comp} for more computational details.}
\label{fig:longbeam:3}
\end{figure}

The proposed approach is also able to control the structure's thickness to meet different device requirements, with addition direct usage of a density filter for thickness control.
Fig.~\ref{fig:longbeam:3} shows the obtained self-supporting structures of thicknesses $1.5$, $2$ and $3$. The overall structures of these three self-supporting structures are similar, and as the minimum thickness increases, the slender beams are removed gradually. Consequently, the smaller the thickness required the more details preserved in the final structures. The associated compliances of the three structures are very close to each other, as summarized in Table~\ref{table:comp}, of a ratio to their reference structures $C/C_{ref}=100.11\%,101.43\%,100.32\%$.

\subsubsection{Different volume fractions}
Performance of the approach is also tested at constraints of different volume fractions, and the computed structures are shown in~\label{fig:irr-res}. Such self-supporting structures become harder to obtain for small value of volume fractions. As can be observed from the results, the self-supporting constraints can still be satisfied although the number of elements decreases as the volume fraction becomes smaller.
\begin{figure}[bt]
\centering
\subfigure[0.6]{\includegraphics[angle=90,width=0.1\textwidth]{long-beam-sup-15.pdf}}
\subfigure[0.5]{\includegraphics[angle=90,width=0.1\textwidth]{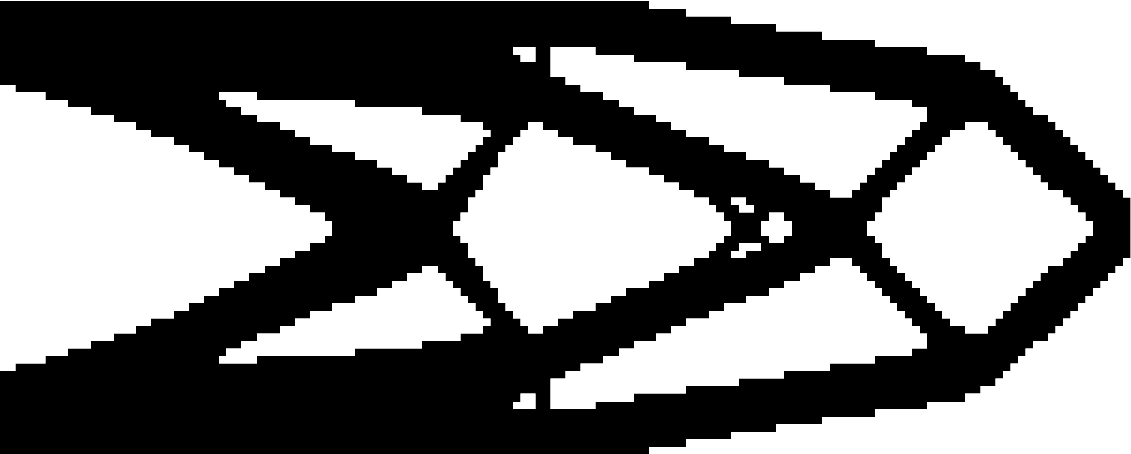}}
\subfigure[0.4]{\includegraphics[angle=90,width=0.1\textwidth]{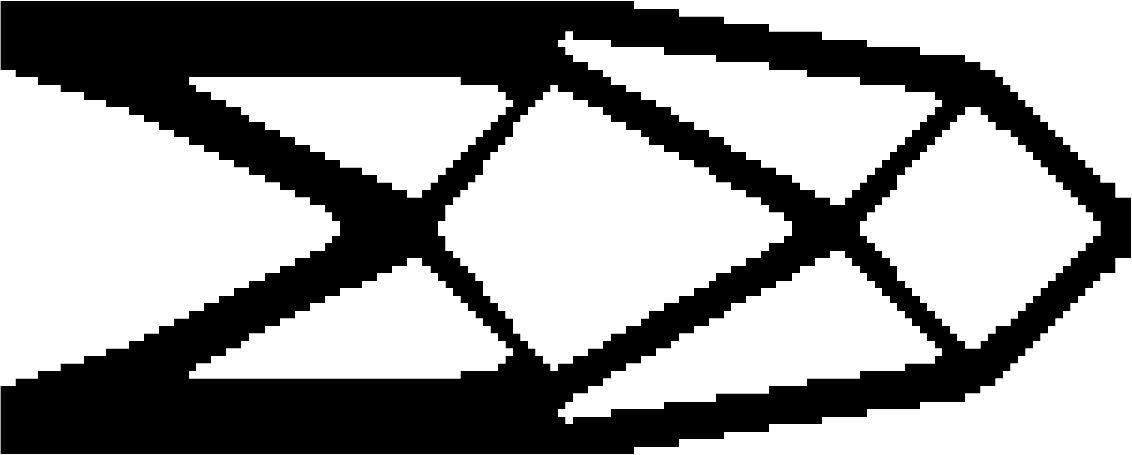}}
\subfigure[0.25]{\includegraphics[angle=90,width=0.1\textwidth]{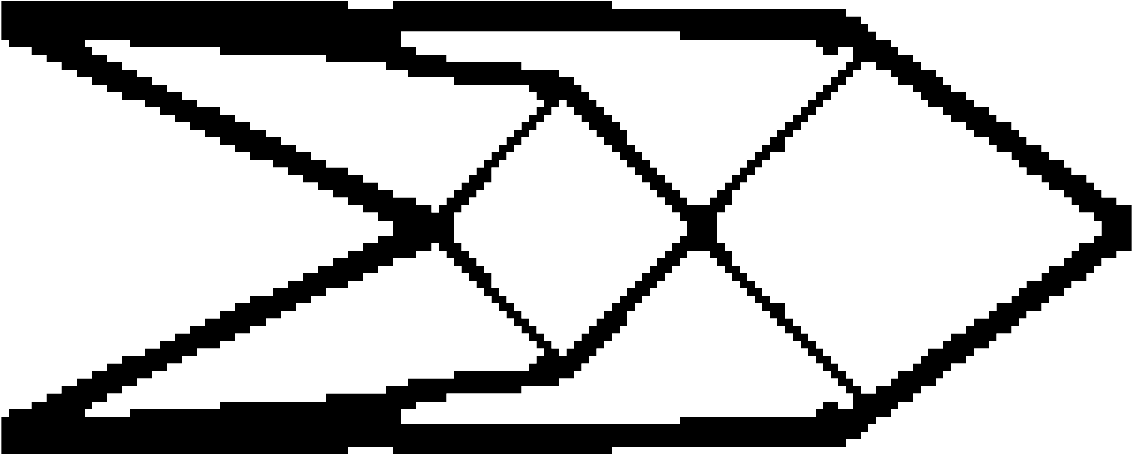}}
\caption{Self-supporting structures computed using the proposed approach at different volume fractions. See also Table~\ref{table:comp} for more computational details.}
\label{fig:irr-res}
\end{figure}

\subsubsection{Different overhang angles}\label{sec-extension}
As have been explained previously in Section~\ref{sec-numerical:angle}, the proposed approach can also work for overhang angle different from $45^{\circ}$ via using different convolution Kernel matrices $\H$. We demonstrate its performance still using the Cantilever beam example in Fig.~\ref{fig:cantilever}(a), for three different overhang angles: $30^{\circ}, 45^{\circ}, 60^{\circ}$. The associated convolution kernel matrices $\H$ for angles of $30^{\circ}, 60^{\circ}$ are also shown in Fig.~\ref{fig:irr-res}(d),(e). As can be observed from the examples, as the overhang angle becomes bigger, the boundary edges moves upward to meet the self-supporting constraints. The structure's minimal compliance is still well kept at these different angles, with $C/C_{ref}$ respectively of $107.37\%$,$100.95\%$ and $133.55\%$. It is also noticed that the larger overhang angle deteriorates the structures' physical performance. Similar phenomenon were also observed in previous studies~\cite{Gaynor2016,Lang20161,Lang20162}.

\begin{figure}[bt]
\centering
\subfigure[$30^{\circ}$]{\includegraphics[angle=90,width=0.1\textwidth]{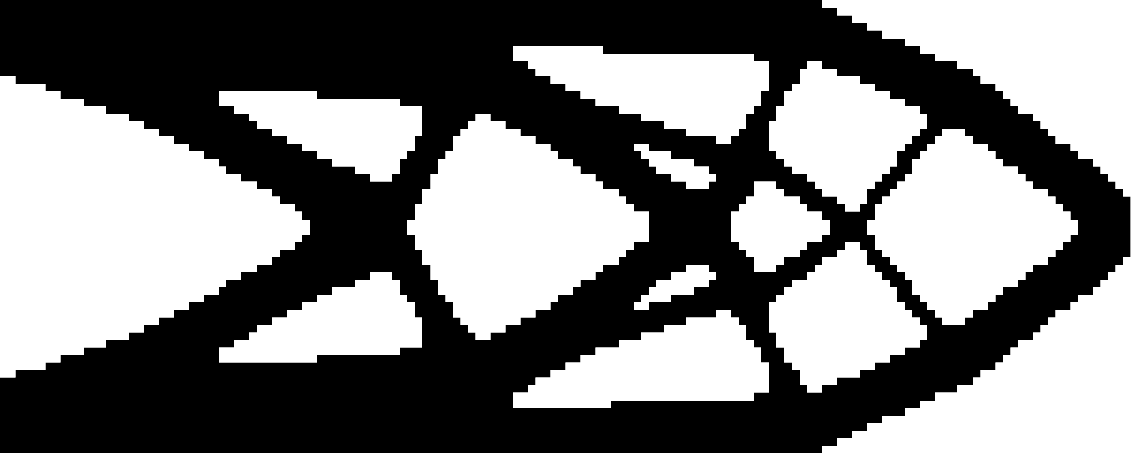}}\quad
\subfigure[$45^{\circ}$]{\includegraphics[angle=90,width=0.1\textwidth]{res45-05.pdf}}\quad
\subfigure[$60^{\circ}$]{\includegraphics[angle=90,width=0.1\textwidth]{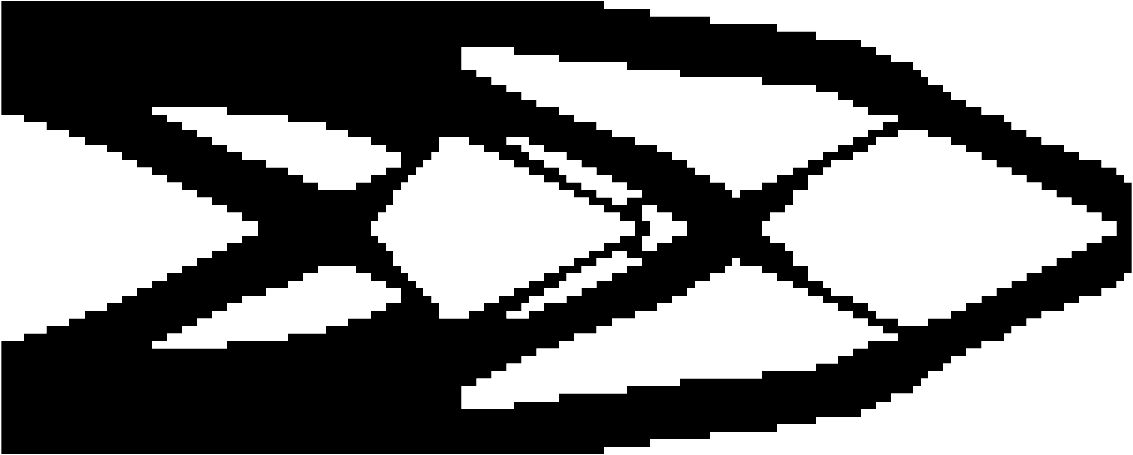}}
\subfigure[Kernel matrix $\H$ at $30^{\circ}$]{\includegraphics[width=0.23\textwidth]{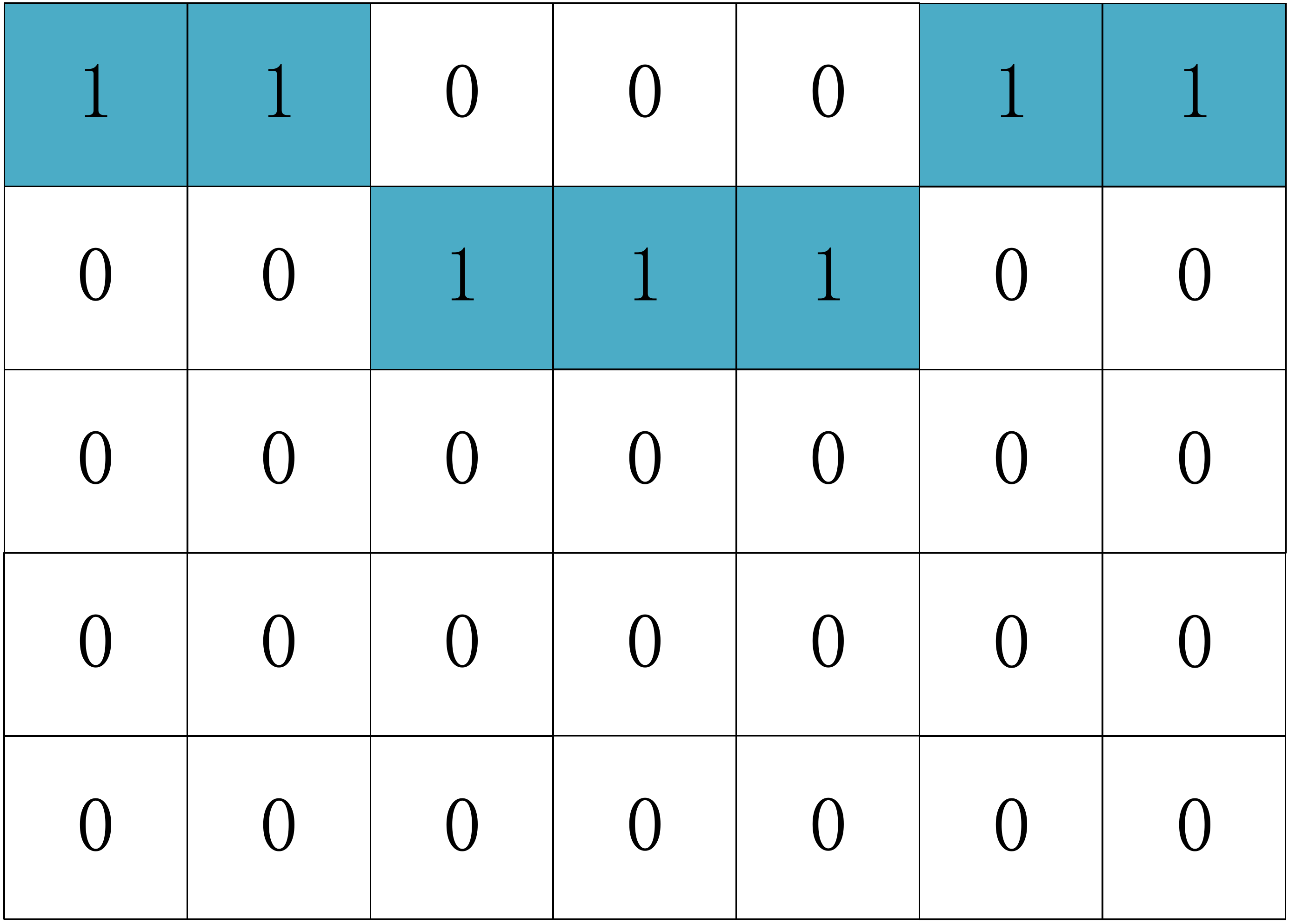}}\quad
\subfigure[Kernel matrix $\H$ at $60^{\circ}$]{\includegraphics[width=0.1\textwidth]{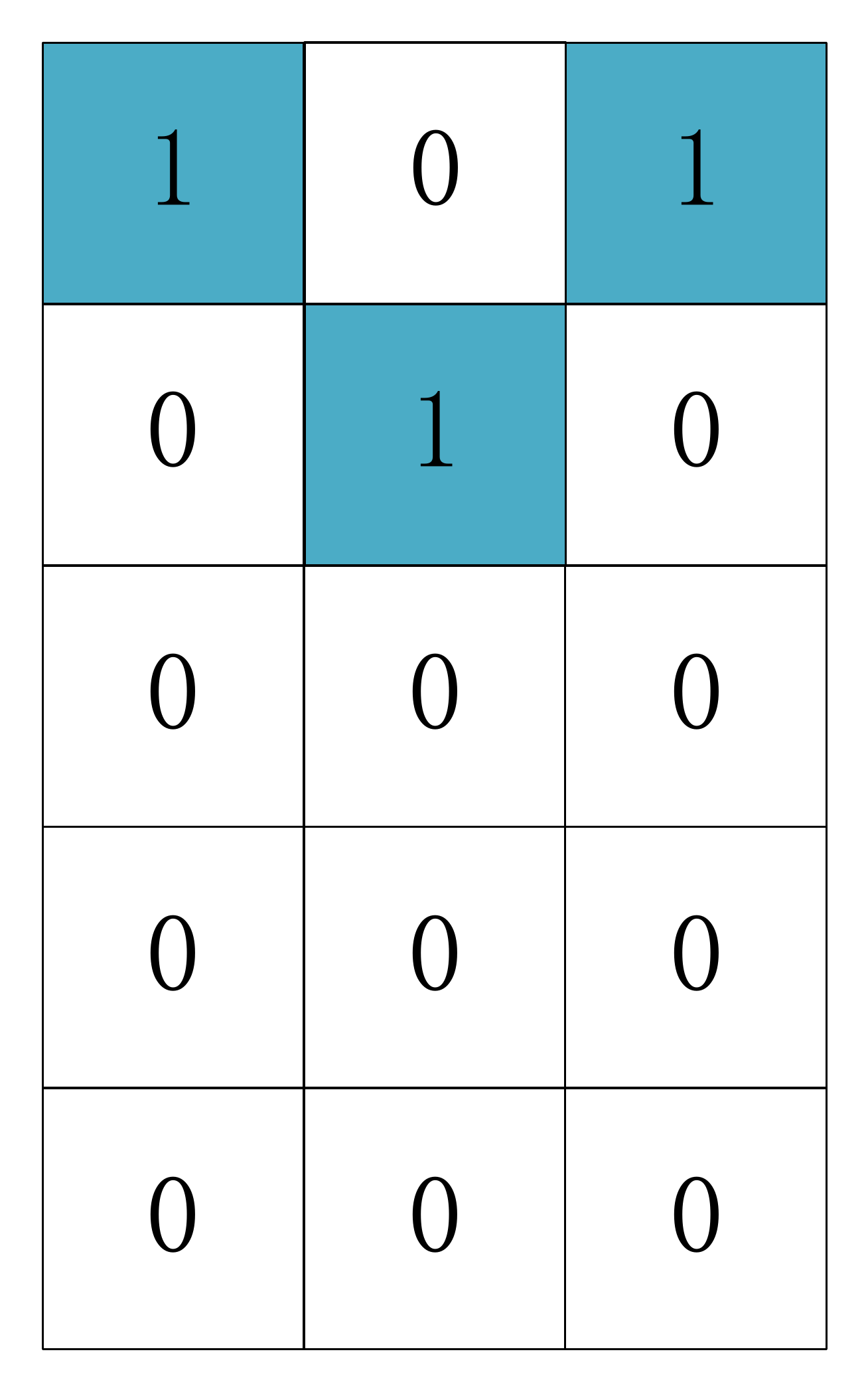}}
\caption{Self-supporting structures obtained using the proposed approach at different overhang angles, and their associated convolution Kernel matrices. See also Table~\ref{table:comp} for more computational details.}
\label{fig:irr-res} 
\end{figure}

\subsection{2D Cantilever beam example with different types of forces}
\begin{figure}[bt]
\centering
\includegraphics[width=0.46\textwidth]{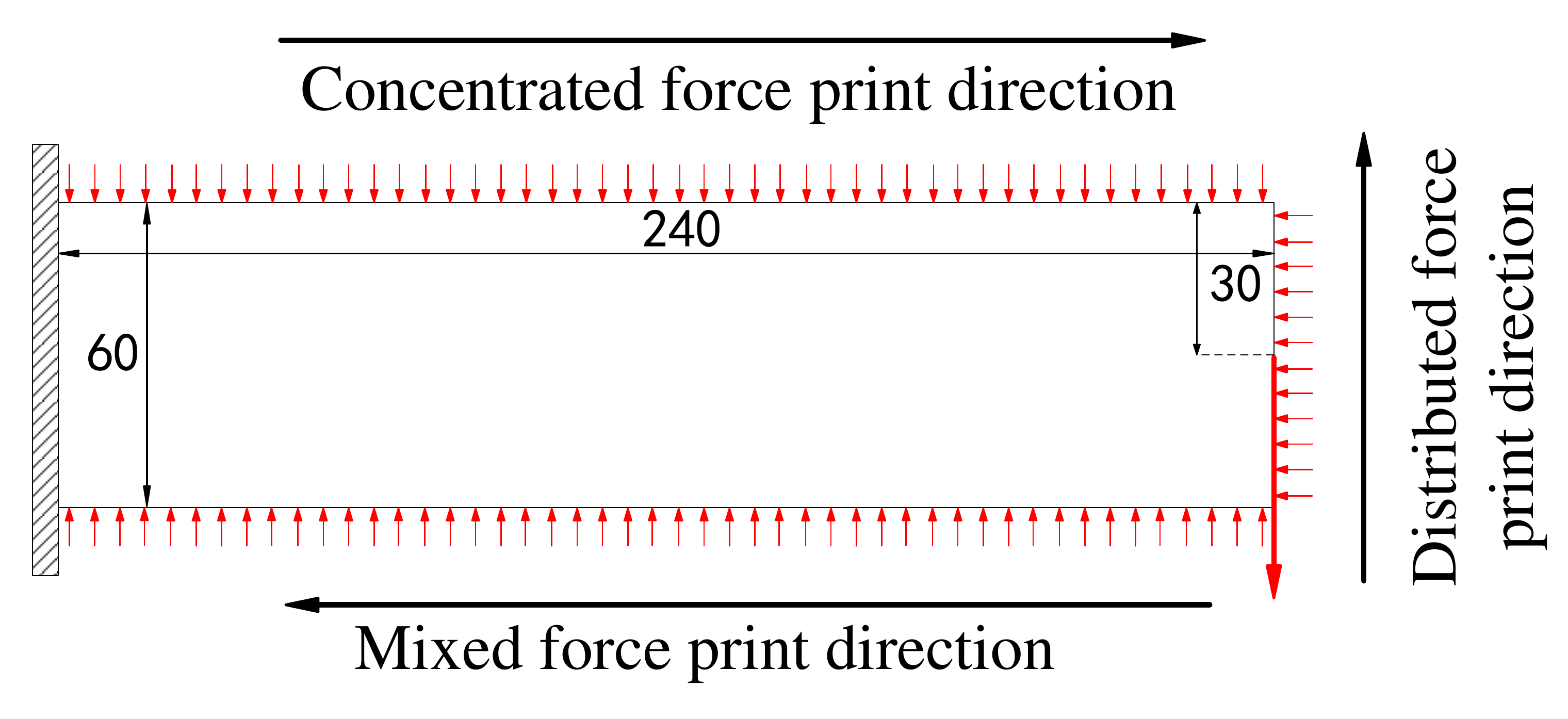}
\caption{The Cantilever beam example with distributed external forces.}
\label{fig:forces}
\end{figure}

In order to test the ability of the proposed approach in selecting the print direction and its performance in finding the optimal structure at different external loadings, the 2D Cantilever beam example in Fig.~\ref{fig:forces} is tested under different types of forces, respectively of concentrated force, distributed force and mixed forces. In this example, the design domain is discretized into $240 \times 60$ square FE elements. The volume fraction is $0.6$ and the minimum thickness is $2$. The concentrated force is exerted on the middle point of the right edge and points downward. The distributed force is exerted evenly on the bottom, top and right edges of the model, while the case of mixed forces takes them both into account.

\begin{figure}[bt]
\centering
\subfigure[Point force] {\includegraphics[angle=90,width=0.08\textwidth]{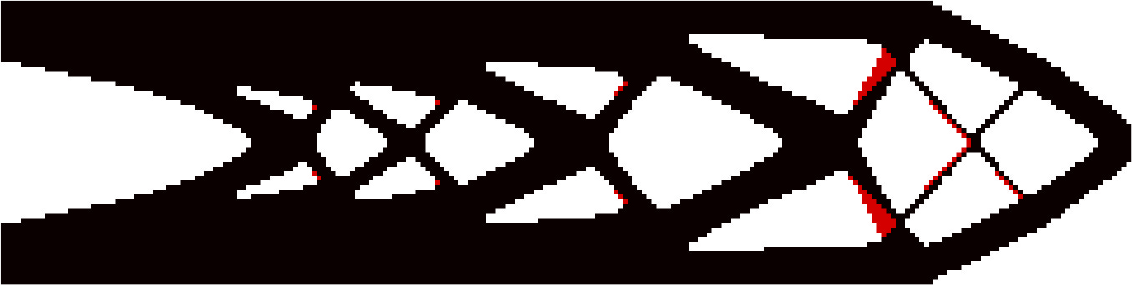}} \quad
\subfigure[Point force] {\includegraphics[angle=90,width=0.08\textwidth]{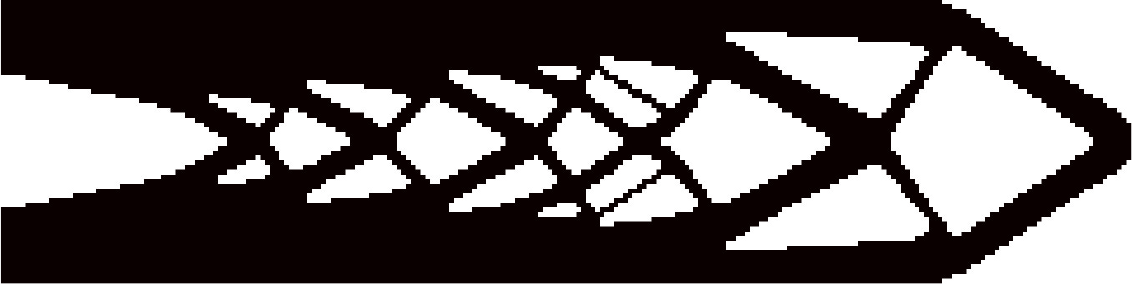}} \quad
\subfigure[Distr. forces] {\includegraphics[angle=90,width=0.08\textwidth]{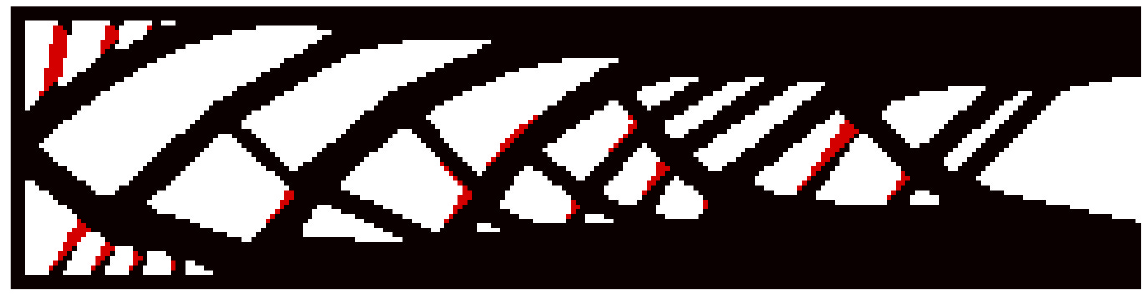}} \quad
\subfigure[Distr. forces] {\includegraphics[angle=90,width=0.08\textwidth]{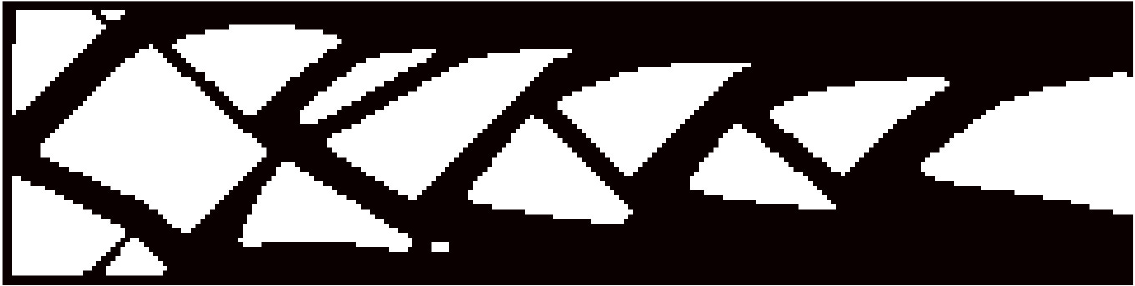}}
\\
\subfigure[Point and distributed forces] {\includegraphics[width=0.45\textwidth]{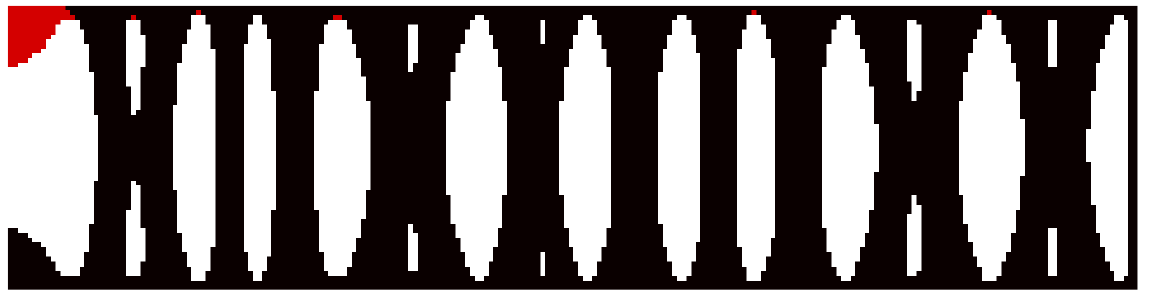}}\quad
\subfigure[Point and distributed forces] {\includegraphics[width=0.45\textwidth]{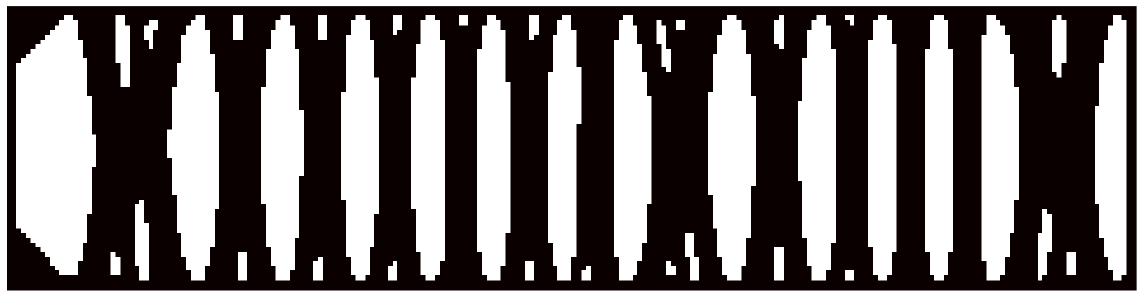}}
\caption{The derived optimal structures of the Cantilever beam example in Fig.~\ref{fig:forces} under different types of external forces. In all cases, structure similarity and compliance closeness is observed between the structures with or without considering the self-supporting constraints. Different print directions were chosen for these different cases for ease of convergence and optimal physical performance. See also Table~\ref{table:comp} for more computational details.}
\label{fig:forces:results}
\end{figure}

For each of the three cases, different print directions were chosen and shown in Fig.~\ref{fig:forces} by the proposed approach. The corresponding optimal structures are also shown in Figs.~\ref{fig:forces:results} respectively. As can be observed from the results and the summary in Table.~\ref{table:comp}, different boundary conditions may require different print directions and produce different optimized structures, which all can be handled successfully via the proposed approach. Compliance of the reference structure are maintained at a compliance ratio respectively of $100.31\%$, $100.27\%$ and $101.70\%$.

\subsection{2D MBB example constrained by more than one print direction}
\begin{figure}[tb]
\centering
\subfigure[Half domain]{\includegraphics[width=0.3\textwidth]{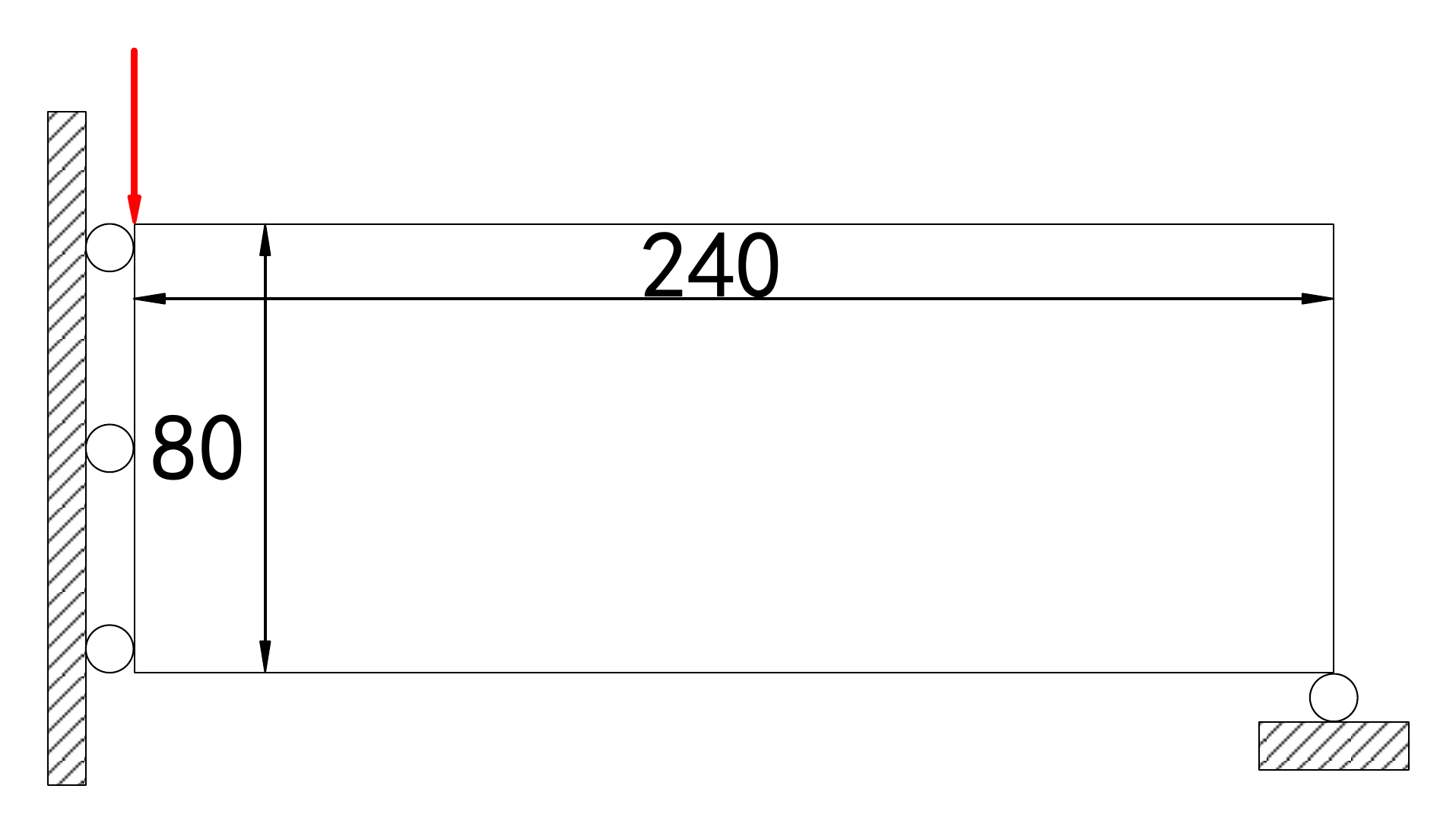}}
\caption{The 2D MBB example: due to its symmetry only one half of the model needs to be studied in the topology optimization process. }
\label{fig:mbb}
%
\centering
\subfigure[Support-needed]{\includegraphics[width=0.48\textwidth]{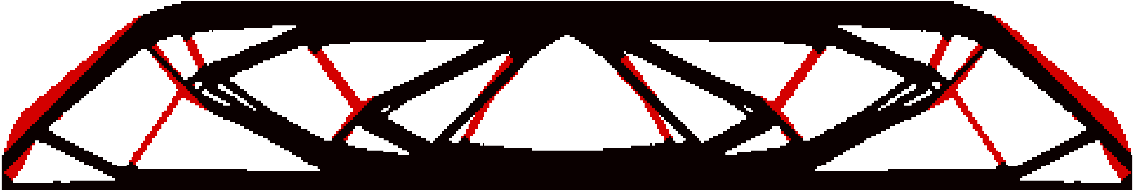}}
\subfigure[Self-supporting]{\includegraphics[width=0.48\textwidth]{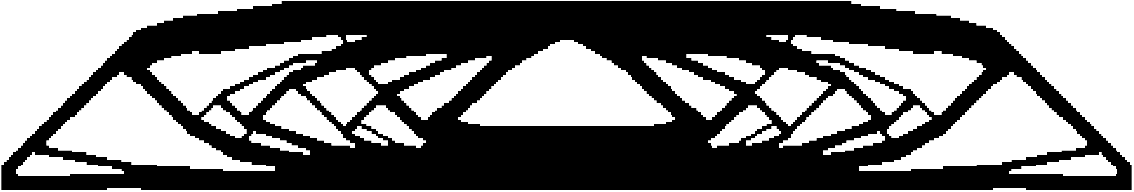}}
\caption{The topology optimization result of the MBB model without (a) or with (b) self-supporting structure. The self-supporting constraint is needed simultaneously in two different directions: from left to right and from right to left, so that the mirror symmetry of the original structure can be maintained. See also Table~\ref{table:comp} for more computational details.}
\label{fig:mbb:result}
\end{figure}

The proposed approach is also able to simultaneously take into account more than one print direction constraints, provided they do not conflict with each other. This is illustrated using the classical MBB problem in Fig.~\ref{fig:mbb}. Due to the symmetry of the model's structure and boundary conditions, only half of the computational domain is used here which consists of $160 \times 30$ square FE mesh elements. The volume fraction is $0.5$ and the minimum thickness is $1.5$.

The aim is to produce a self-supporting structure maintaining the mirror symmetry of the original model. Thus the self-supporting requirement has to be added in both directions: from right to left and from left to right for the half sized structure in Fig.~\ref{fig:mbb}(b). As a result, a self-supporting in both directions is obtained in Fig.~\ref{fig:mbb:result}, as compared with the support-needed structure in Fig.~\ref{fig:mbb:result}(a), where the unsupported elements are plotted in red. The compliance of the support-needed structure and the self-supporting structure are respectively 185.7 and 191.3, at a ratio of $103.02\%$.

Note that the 2D MBB problem of the same domain size was also tested by Gaynor and Guest in~\cite{Gaynor2016}, where the print direction was manually set from bottom to top. Such setting thus does not require constraints of multiple constraints.

\subsection{Complex internal structure at distributed force}
\begin{figure*}[bt]
\centering
\subfigure[Problem]{\includegraphics[width=0.22\textwidth]{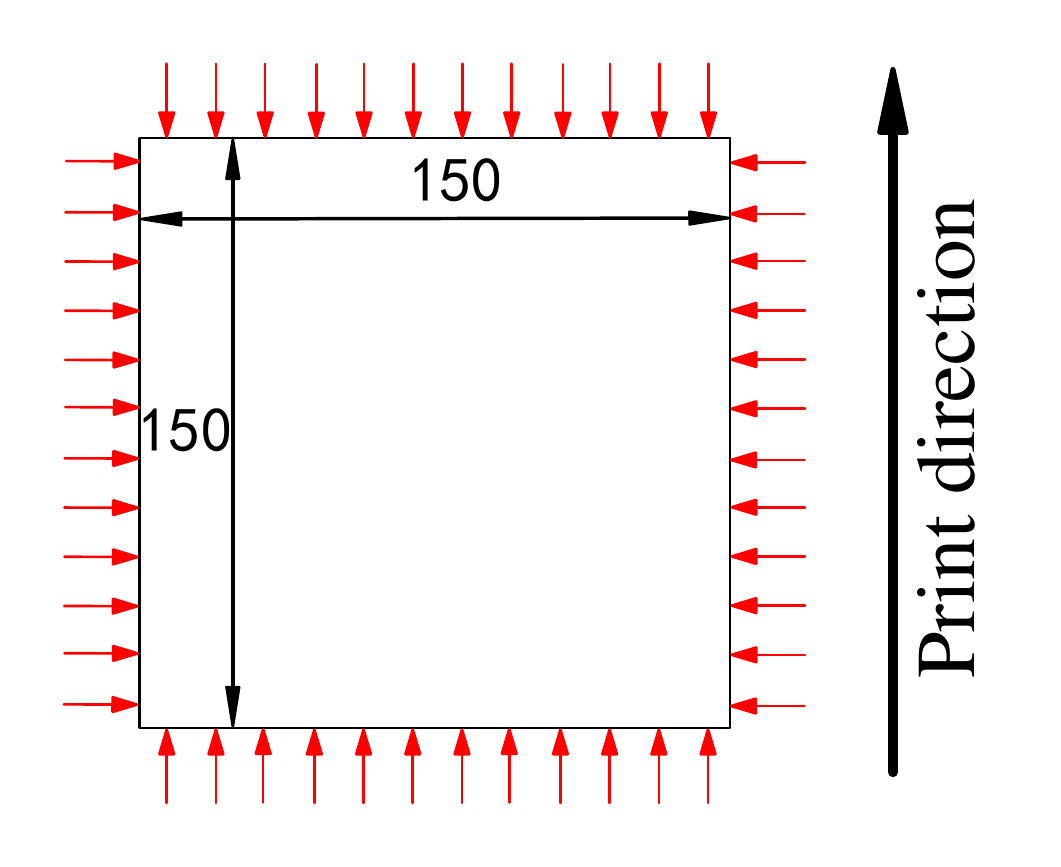}}\qquad
\subfigure[Support-needed]{\includegraphics[width=0.2\textwidth]{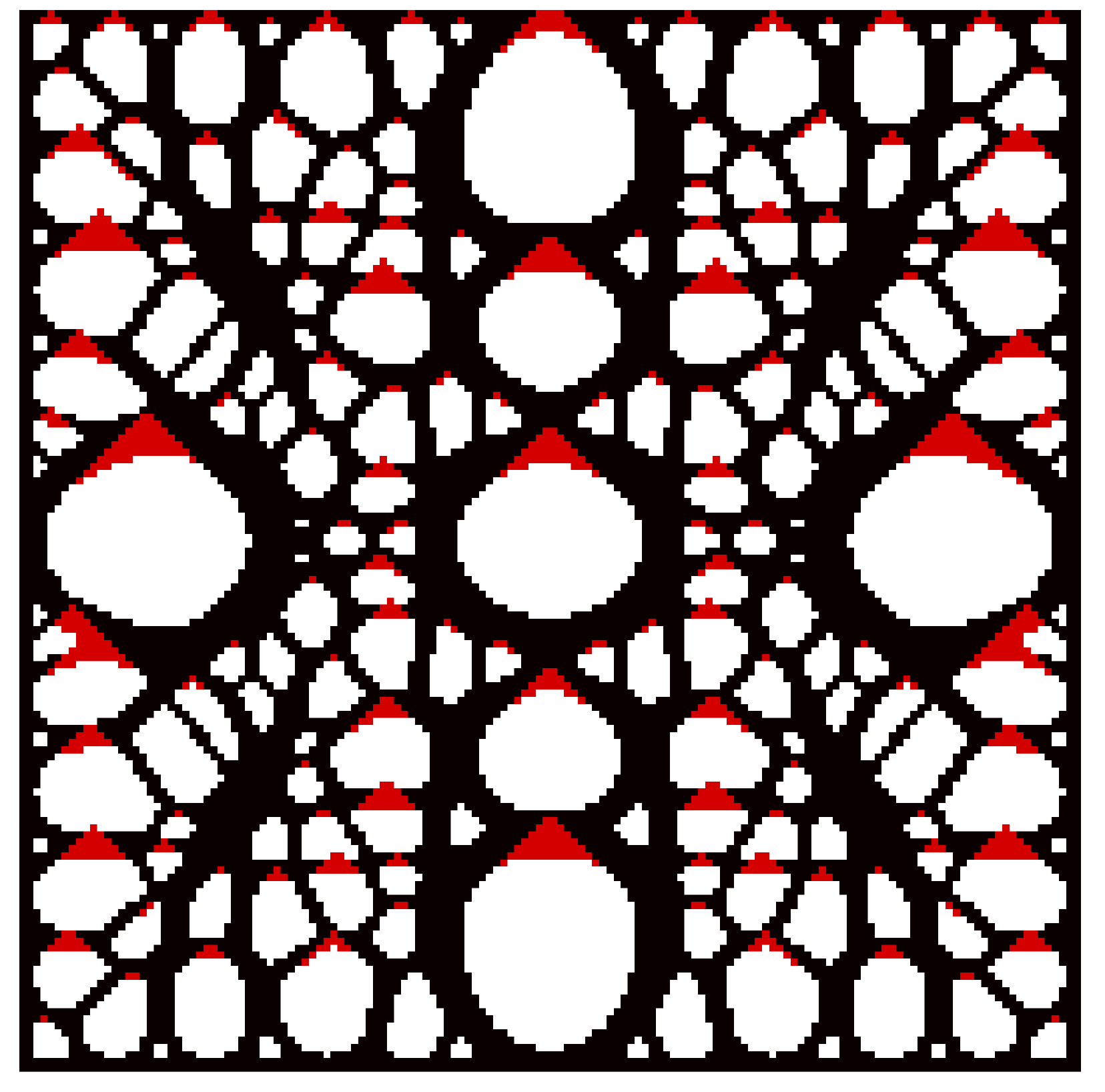}}\quad
\subfigure[Self-supporting]{\includegraphics[width=0.2\textwidth]{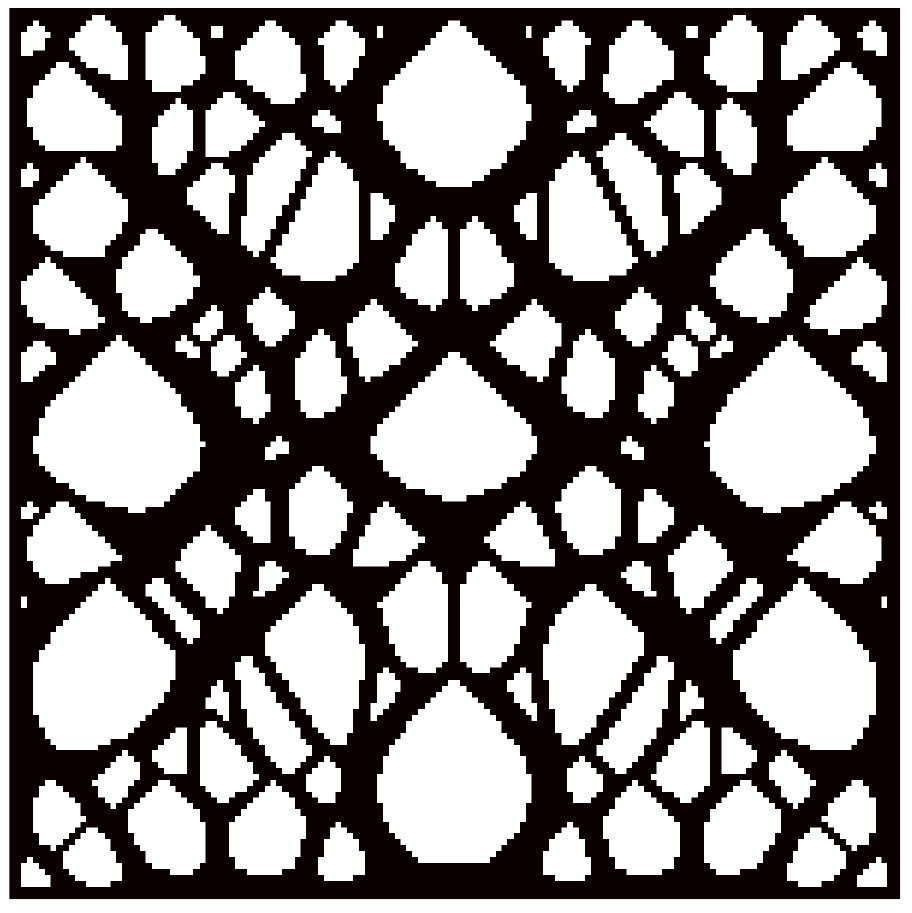}}
\caption{The problem of a square under a distributed force, and its computational results. See also Table~\ref{table:comp} for more computational details.}
\label{fig:square}
\end{figure*}

The proposed approach is also able to produce self-supporting structure for complex internal structure, as demonstrated using the square example at distributed forces in Fig.~\ref{fig:square}(a). The computed support-needed structure and self-supporting structure are respectively given in Fig.~\ref{fig:square}(b) and (c). It can be seen from the results that the original support-needed structure has many small flat edges which prevent the structure to be fabricated without a large number of additional supports. Such unsupported elements have disappeared in the optimized self-supporting structure of Fig.~\ref{fig:square}(b), despite the structure's high complexity. In addition, the self-supporting structure also has a close compliance to that of the original support-needed structure, respectively of $1544.7$ and $1312.8$ at a compliance ratio of $117.66\%$. A self-supporting structure is very necessary for such highly complex shape, as computing the supports or its removal would be extremely troublesome if not impossible.


\subsection{3D examples}
The proposed approach allows for a parallel implementation, which is particularly important for complex 3D examples of high DOFs. We have implemented the approach in GPU for parallel computations, and it works efficiently for examples of millions of elements within almost an hour. The test examples includes the classical benchmark examples: a 3D wheel, a 3D Cantilever, and a newly devised examples of a 3D desk. The domain sizes and their associated computational time is summarized in Table~\ref{table:time}.
\begin{table*}[tb]
\centering\small
\caption{\label{table:time} Time comparison for 3D examples}
\begin{tabular}{lccccccccc}
\toprule
&Example  & Wheel in Fig.~\ref{fig:wheel} & Cantilever(a) in Fig.~\ref{fig:wheel} & Cantilever(b) in Fig.~\ref{fig:wheel} & Desk in Fig.~\ref{fig:desk}p \\
\midrule
&Size  & $100\times 100\times 100$ & $100\times 50\times 50$ & $150\times 50\times 50$ & $120\times 120\times 80$ \\
&Time  & 35m & 11m & 17m & 66m \\
\bottomrule
\end{tabular}
\end{table*}

\subsubsection{3D Wheel}
\begin{figure}[bt]
\centering
\includegraphics[width=0.3\textwidth]{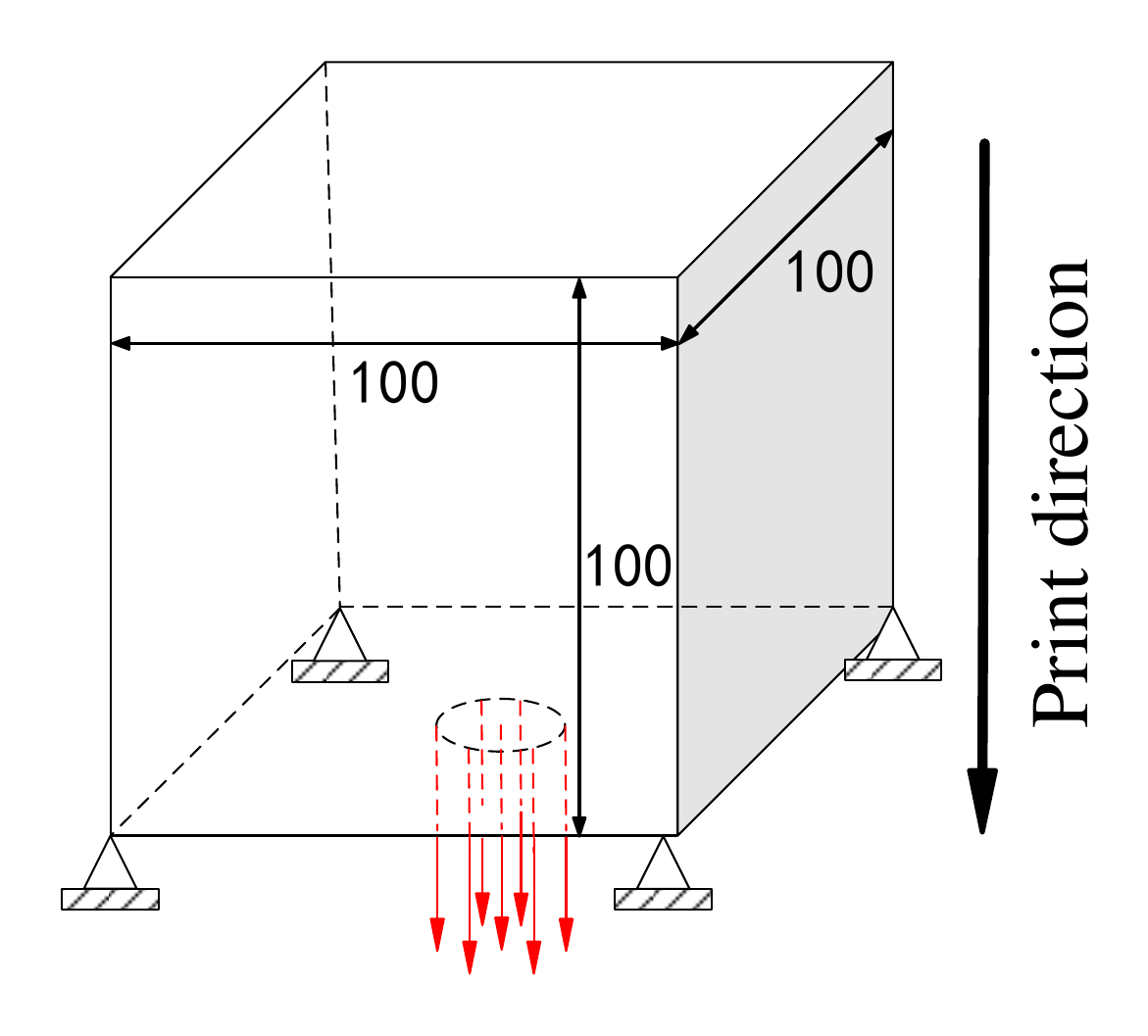}
\caption{The wheel example.}
\label{fig:wheel}
\end{figure}
\begin{figure}[bth]
\centering
\subfigure[Support-needed]{\includegraphics[width=0.24\textwidth]{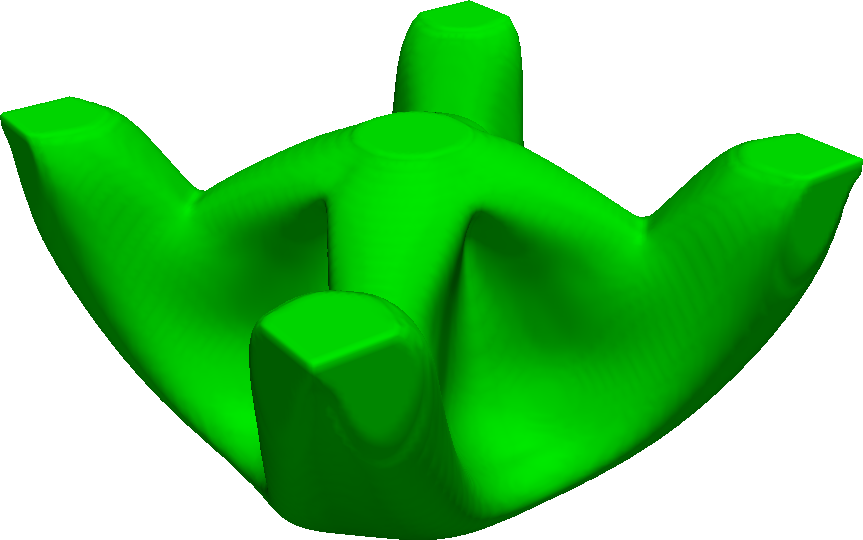}}
\subfigure[Self-supporting]{\includegraphics[width=0.23\textwidth]{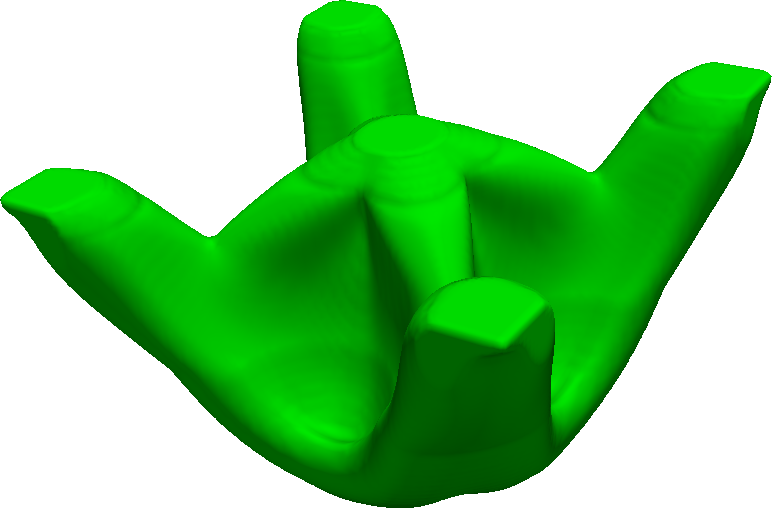}}
\subfigure[Support-needed, $x=25$]{\includegraphics[width=0.15\textwidth]{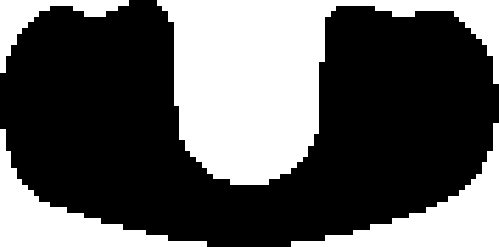}}
\subfigure[Support-needed, $x=35$]{\includegraphics[width=0.15\textwidth]{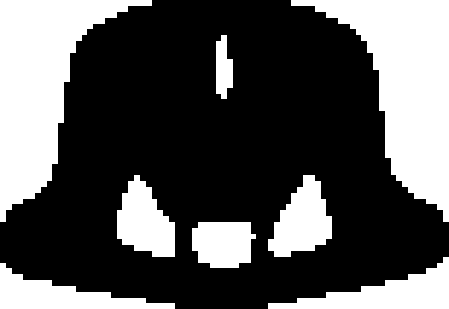}}
\subfigure[Support-needed, $x=45$]{\includegraphics[width=0.15\textwidth]{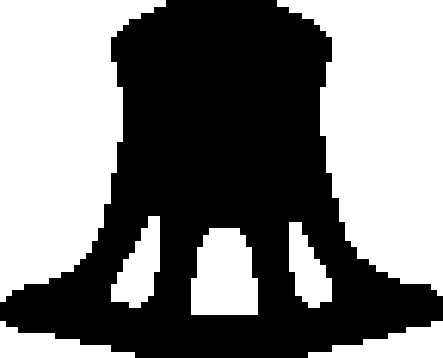}}
\subfigure[Self-supporting, $x=25$]{\includegraphics[width=0.15\textwidth]{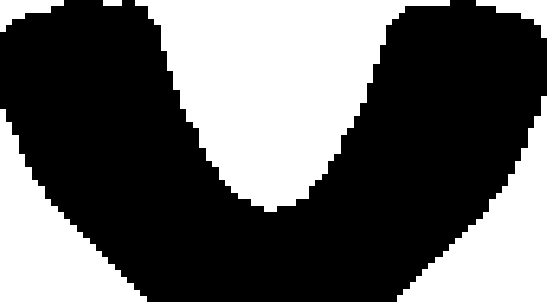}}
\subfigure[Self-supporting, $x=35$]{\includegraphics[width=0.15\textwidth]{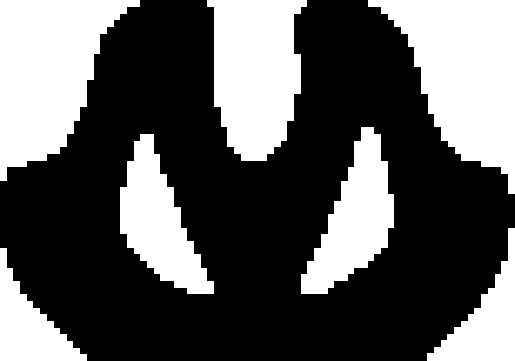}}
\subfigure[Self-supporting, $x=45$]{\includegraphics[width=0.15\textwidth]{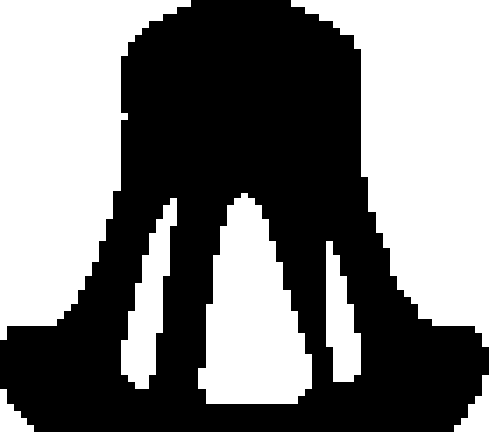}}
\caption{The self-supporting structure generated using the proposed approach and its comparison with that of the support-needed one generated without the consideration of self-supporting constraint. Both the 3D structure and their slices at $x=25,35,45$ are shown.}
\label{fig:wheel:results}
\end{figure}

The 3D wheel example in Fig.~\ref{fig:wheel} consists of $100 \times 100 \times 100$ cubic mesh elements. The four corners of the bottom face are fixed and a concentrated force is exerted on the middle point of the bottom face. The target volume fraction is $0.25$ and the minimum thickness is $2$. The print direction is chosen as top to bottom.

The computed self-supporting structure using the proposed approach is shown in Fig.~\ref{fig:wheel:results}(b), as compared with its counterpart of support-needed structure in Fig.~\ref{fig:wheel:results}(a). The corresponding structure slices at $x=25,35,45$ of both the self-supporting and support-needed structures are also shown and compared in Fig.~\ref{fig:wheel:results}(f)(g)(h). As can be seen from the results, the originally flat regions of the support-needed structure, which cannot be fabricated without supports, have been optimized to meet the self-supporting requirement. The resulting structure is totally self-supporting for direct fabrication purpose and its compliance is $87.32$, very close to that of the original support-needed one of $86.16$, of a ratio $1.01\%$

\subsubsection{3D Cantilever}
Two different 3D Cantilever examples are tested here as illustrated in Fig.~\ref{fig:cantilever}: one of size $100 \times 50 \times 50$ exerted by point loadings, and another one of size $150 \times 50 \times 50$ exerted by edge loadings; a same example to the latter was also studied in~\cite{Lang20162}. The target volume fraction is $0.3$ and the minimum thickness is $2$. The print direction is chosen as top to bottom. The computed self-supporting structures are shown in Fig.~\ref{fig:cantilever}, which is directly to be fabricated without any additional support materials.

\begin{figure}[bt]
\centering
\includegraphics[width=0.3\textwidth]{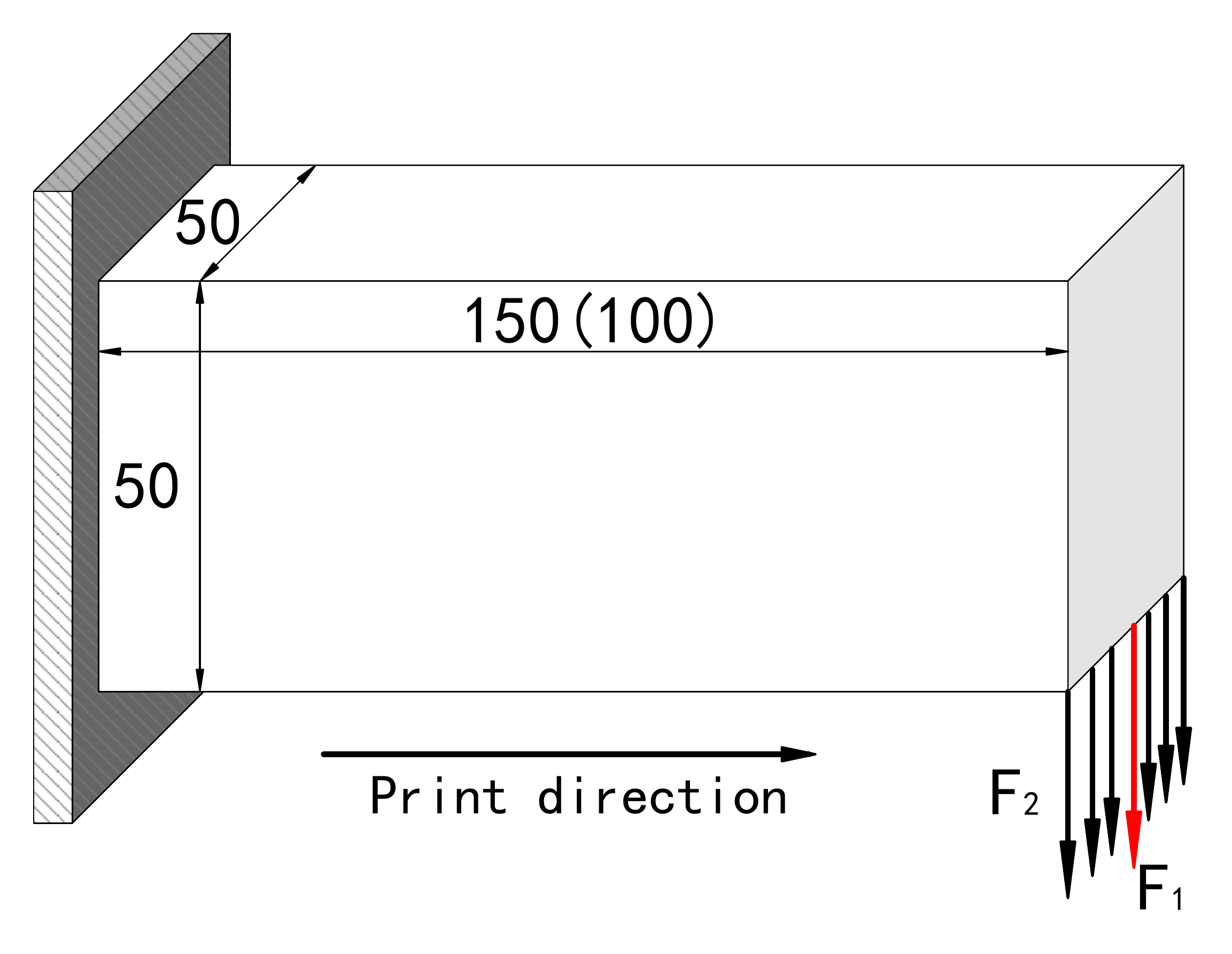}
\caption{A 3D Cantilever example of size $100\times 50\times 50$ at a point loading $F_1$, and of size $150\times 50\times 50$ at an edge loading $F_2$.}
\label{fig:3dbeam}
\end{figure}
\begin{figure}[bt]
\centering
\subfigure[Self-supporting]{\includegraphics[width=0.15\textwidth]{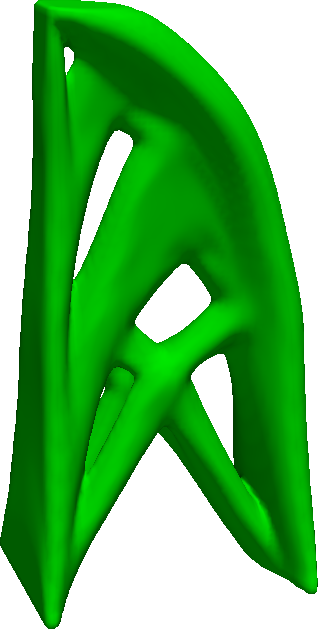}}\quad
\subfigure[Slice in $x=25$]{\includegraphics[width=0.15\textwidth]{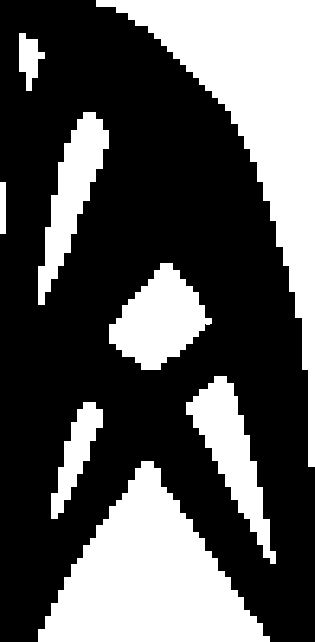}}\\
\subfigure[Self-supporting]{\includegraphics[width=0.135\textwidth]{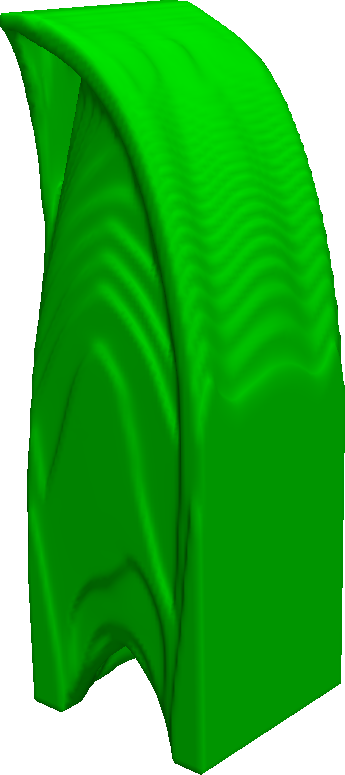}}
\subfigure[Self-supporting]{\includegraphics[width=0.12\textwidth]{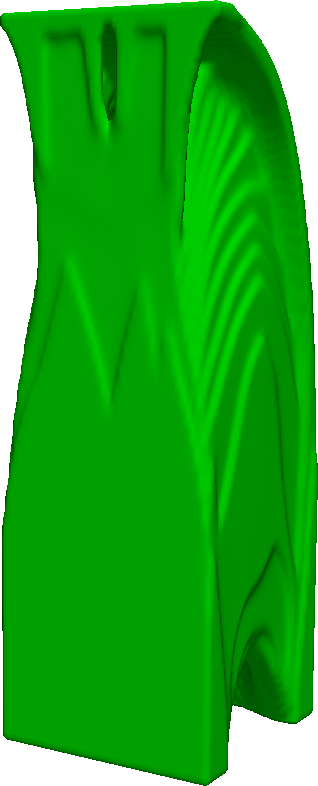}}
\subfigure[Slice in $x=15$]{\includegraphics[width=0.1\textwidth]{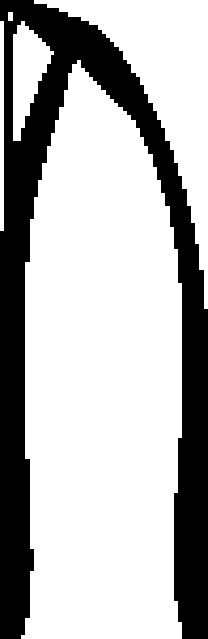}}
\subfigure[Slice in $x=20$]{\includegraphics[width=0.1\textwidth]{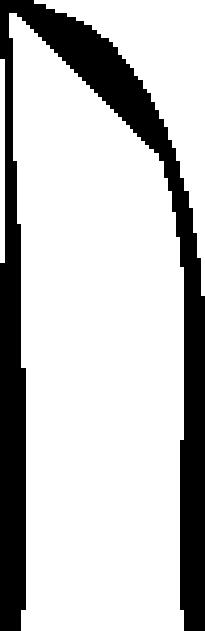}}
\caption{The generated self-supporting structures, together with its slice, for two difference cases: (a),(b) for structure of size $100\times 50\times 50$ at a point loading $F_1$, and (c),(d),(e),(f) of size $150\times 50\times 50$ at an edge loading $F_2$ for the example in Fig.~\ref{fig:3dbeam}.}
\label{fig:cantilever}
\end{figure}

\subsubsection{3D Desk}
\begin{figure}[bt]
\centering
\includegraphics[width=0.3\textwidth]{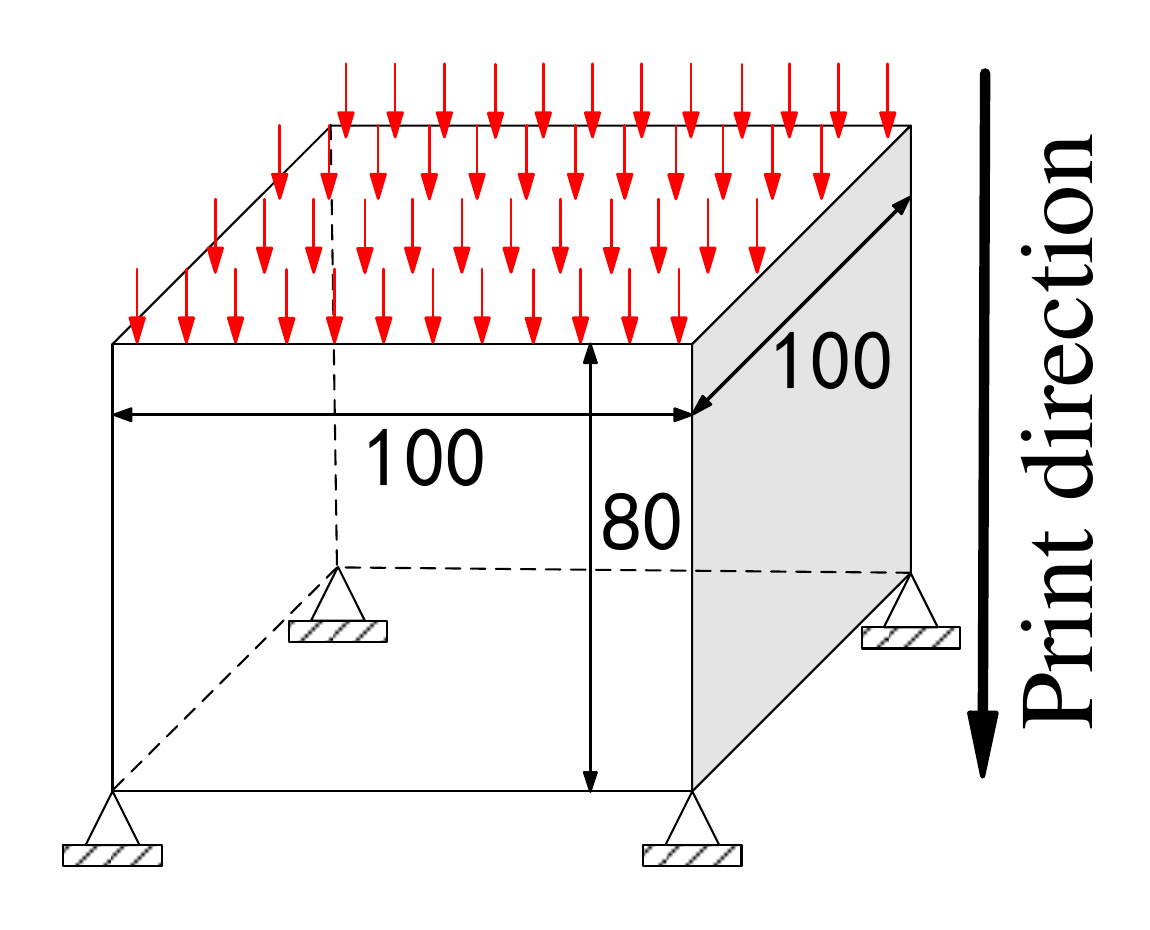}
\caption{A 3D desk example.}\label{fig:desk}
\end{figure}

A more complex 3D desk problem, as shown in Fig.~\ref{fig:desk}, is designed to further test performance of the proposed approach. The example is of size $120 \times 120 \times 80$, and consist of $1.152$ millions of elements. In this example, the four bottom corners are fixed and the top face is exerted by a uniform distributed force pointing downward. The target volume fraction is $0.3$, the minimum thickness is $2$ and the print direction is chosen from right to left. The final generated self-supporting structure using the proposed approach is shown in Fig.~\ref{fig:desk}, taking $66$ minutes. It is also interesting to note that the four legs of the desk are not totally solid but take a porous bone-like structure to balance the constraint of the object weight and the target compliance.

\begin{figure}[bt]
\centering
\subfigure[Self-supporting structure]{\includegraphics[width=0.3\textwidth]{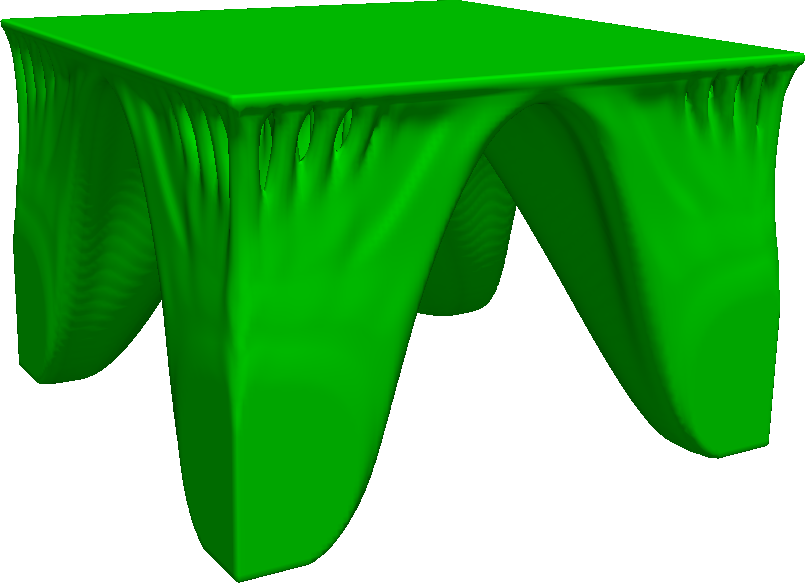}}\\
\subfigure[Slice in $x=25$]{\includegraphics[width=0.23\textwidth]{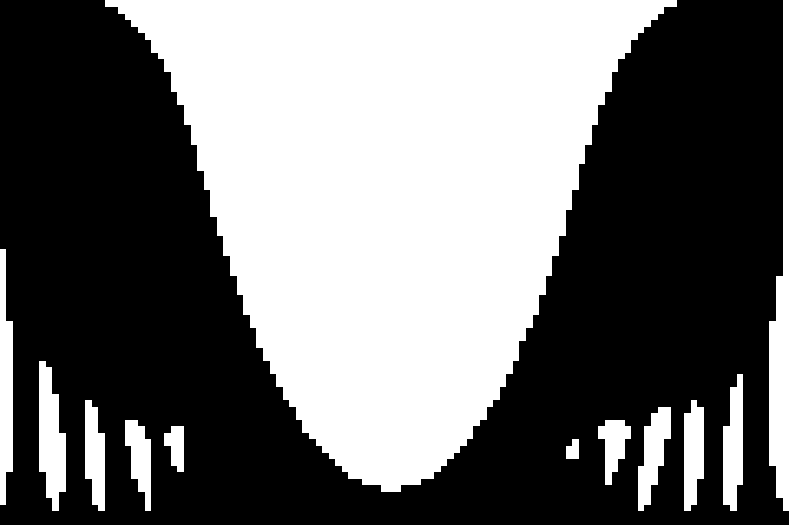}}
\subfigure[Slice in $x=35$]{\includegraphics[width=0.23\textwidth]{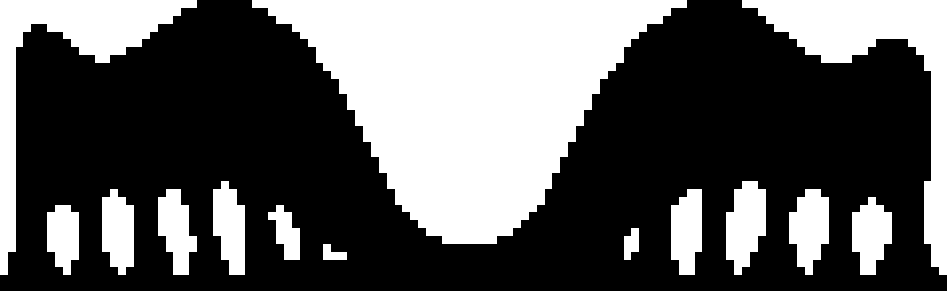}}
\caption{The generated self-supporting structure and the associated slices at $x=23$ and $x=35$, for the 3D desk example in Fig.~\ref{fig:desk}.}
\label{fig:desk:results} 
\end{figure}

\subsection{Summary}
As can be observed from these examples, using the proposed self-supporting topology optimization approach, a completely self-supporting structure is achieved that has a very similar geometric shape to that of the support-needed one without considering the self-supporting constraint. The edges or faces of the original support-needed structure, are alighted toward the print direction so that all elements can be successfully fabricated. In addition, the difference in compliance between the self-supporting structure and the support-needed structure is maintained within a very small or negligible range. All these enhancements demonstrate the strength of the proposed approach in designing the self-supporting structures and simultaneously maintaining their optimized structure and physical performance.

\section{Conclusion}\label{sec-conclusion}
A novel self-supporting topology optimization approach is provided in this paper applicable to additive manufacturing. The usage of convolution operator and the associated numerical techniques enables the self-supporting structure to be reliably generated with high efficiency and robustness. Extensive 2D and 3D examples are provided to test the approach's performance. The final derived structures are completely self-supporting and have a compliance very close to the optimal support-needed structure, proving its high effectiveness.

The proposed approach is at present implemented using regular square or cubic elements, which are dominated in researches of topology optimization. On the other hand, the overall framework also works for general domains consisting of irregular quad- or hex- elements, but is also limited by the fact that the convolution operator presented in Section~\ref{sec-numerical:conv} is no longer applicable as they become different for different elements. Thus the element by element enumeration has to be taken, and will reduce the computational efficiency.

The proposed approach chooses an appropriate printing direction before taking into account the self-supporting constraint during the optimization process. The strategy aims to provide a good initial value to improve the convergence and to result in a possible smaller target compliance. On other hand, if ignoring the step of choosing the print angle and setting an arbitrary print direction, the approach may fail to produce a converged result. For example, an optimized self-supporting structure is hard to obtain for the MBB example at a different direction. Such phenomenon may not prevent the approach in generating a self-supporting structure suitable for additive manufacturing, but may hinder its usage for specific applications. It deserves further researches efforts to improve this.


The proposed approach can also be extended to porous interior designs of 3D free-form structures that do not need any additional supports within its interior. Such supports would otherwise be very difficult to remove. This novelty will be very useful in some additive manufacturing technologies and is currently an active area of research. Furthermore, besides the self-supporting requirements, other constraints such as hanging bridge also need to be considered in the optimization process so that a designed structure can be robustly fabricated.

\section*{Acknowledgements}
The work described in this paper is partially supported by the NSF of China (No. 61472356, 61210007).

\section*{References}
\bibliography{support_free}
\end{document}